\documentstyle[psfig]{l-aa}
\begin{document}
\renewcommand{\topfraction}{1.}
\renewcommand{\bottomfraction}{1.}
\renewcommand{\textfraction}{0.}
\thesaurus{06(08.03.1; 08.03.4; 08.13.2; 08.16.4; 11.13.1; 13.09.6)}
\title{Obscured Asymptotic Giant Branch stars in the Magellanic Clouds
       IV. Carbon stars and OH/IR stars\thanks{based on
       observations obtained at the European Southern Observatory (La
       Silla, Chile: proposal ESO 54.E-0135), the South African
       Astronomical Observatory, and the Australia Telescope National
       Facility}}
\author{Jacco Th. van Loon\inst{1,2}, Albert A. Zijlstra\inst{1},
        Patricia A. Whitelock\inst{3}, Peter te Lintel Hekkert\inst{4},
        Jessica M. Chapman\inst{5,6}, Cecile Loup\inst{7,8}, M.A.T.
        Groenewegen\inst{9}, L.B.F.M. Waters\inst{2,10} \and Norman R.
        Trams\inst{11}}
\institute{European Southern Observatory, Karl-Schwarzschild
           Stra{\ss}e 2, D-85748 Garching bei M\"{u}nchen, Germany
      \and Astronomical Institute, University of Amsterdam, Kruislaan
           403, NL-1098 SJ Amsterdam, The Netherlands
      \and South African Astronomical Observatory, P.O.Box 9, 7935
           Observatory, Republic of South Africa
      \and Australia Telescope National Facility, Parkes Observatory,
           P.O.Box 276, Parkes, NSW 2870, Australia
      \and Anglo-Australian Observatory, P.O.Box 296, Epping, NSW 2121,
           Australia
      \and Australia Telescope National Facility, P.O.Box 76, Epping,
           NSW 2121, Australia
      \and European Southern Observatory, Casilla 19001, Santiago 19,
           Chile
      \and Institut d'Astrophysique de Paris, 98bis Boulevard Arago,
           F-75014 Paris, France
      \and Max-Planck Institut f\"{u}r Astrophysik, Karl-Schwarzschild
           Stra{\ss}e 1, D-85740 Garching bei M\"{u}nchen, Germany
      \and Space Research Organization Netherlands, Landleven 12,
           NL-9700 AV Groningen, The Netherlands
      \and ISO Science Operations Centre, Astrophysics Division of
           ESA, Villafranca del Castillo, P.O.Box 50727, E-28080
           Madrid, Spain}
\date{Received date; accepted date}
\maketitle
\markboth{Jacco Th.\ van Loon et al.: Obscured AGB stars in the
          Magellanic Clouds IV}{Jacco Th.\ van Loon et al.: Obscured
          AGB stars in the Magellanic Clouds IV}
\begin{abstract}

We present $N$-band photometry for a sample of 21 dust-enshrouded AGB
stars in the Large Magellanic Cloud, and three additional sources in
the Small Magellanic Cloud. Together with near-infrared photometry,
this is used to give a tentative classification into carbon and
oxygen-rich atmospheres. Bolometric luminosities are also estimated
for these stars. In addition, we present the results of a survey for 
OH masers in the LMC, which resulted in the discovery of OH maser
emission from IRAS04407--7000. Spectra between 600 and 1000~nm have
been obtained for two heavily obscured AGB stars in the LMC,
confirming them to be highly reddened very late M-type giants. Because
the dust-enshrouded stars are clearly undergoing heavy mass loss they
are assumed to be very near the termination of their respective
Asymptotic Giant Branch phases. The fraction of mass-losing
carbon stars decreases with increasing luminosity, as expected from 
Hot Bottom Burning. The best candidate carbon star, with $M_{\rm bol}
\sim -6.8$ mag, is the most luminous mass-losing carbon star in the
Magellanic Clouds, and amongst the most luminous AGB stars. At lower
luminosities ($M_{\rm bol} \sim -5$ mag) both oxygen and carbon stars
are found. This may be explained by a range in metallicity of the
individual mass-losing AGB stars.

\keywords{Stars: carbon -- circumstellar matter -- Stars: mass loss --
Stars: AGB and post-AGB -- Magellanic Clouds -- Infrared: stars}
\end{abstract}

\section{Introduction}

Luminous Asymptotic Giant Branch (AGB) stars in the Large Magellanic
Cloud (LMC) were expected to become carbon stars as a result of third
dredge-up of carbon to the stellar photosphere. The absence of carbon
stars brighter than $M_{\rm bol} \sim -6$ mag came therefore as a big
surprise (Iben 1981). Hot Bottom Burning (HBB: cf.\ Iben \& Renzini
1983) has been proposed as a mechanism to avoid producing luminous
carbon stars by burning the carbon into nitrogen and oxygen before it
reaches the stellar photosphere. Both third dredge-up and HBB are
poorly understood phenomena, and better observational constraints on
the theoretical models are required. It has been stated that in the
LMC there is not only a deficiency of luminous carbon-stars, but a
general deficit of AGB stars more luminous than $M_{\rm bol}\sim-6$
mag: where a few hundred are expected, the observed number is a
factor ten smaller (Frogel et al.\ 1990; Reid et al.\ 1990).

Until recently, searches for AGB stars in the MCs had been limited to
optically bright stars (e.g.\ Blanco et al.\ 1980; Westerlund et al.\
1981; Costa \& Frogel 1996 and references therein). Such stars may
evolve further along the AGB, changing luminosity, chemical
composition and other (circum-)stellar parameters. On the upper AGB
they experience heavy mass loss and become enshrouded in dust, making
them practically invisible at optical wavelengths and accessible only
in the infrared (IR) (see Habing 1996 for a review). These obscured
AGB stars presumably represent the end product of AGB evolution, and
can be used directly to test the predictions of stellar evolution
theories. They may account for some fraction of the missing luminous
AGB stars. The luminous carbon stars, that are absent in samples of
optically visible AGB stars, might also be found amongst the obscured
AGB stars.

After the IRAS satellite opened the thermal-IR window towards the
MCs, the first samples of obscured AGB stars and red supergiants
(RSGs) in the MCs were compiled (Whitelock et al.\ 1989; Reid 1991;
Wood et al.\ 1992). We have significantly extended the sample of
known obscured AGB stars in the MCs. In paper~I (Loup et al.\ 1997)
we selected IRAS point sources as candidate obscured AGB stars in the
LMC. In paper~II (Zijlstra et al.\ 1996) and paper~III (van Loon et
al.\ 1997) new near-IR (NIR) counterparts for a large subsample of
these candidates were discussed. A total of 46 obscured AGB stars in
the LMC and 5 in the SMC have now been identified, allowing a
detailed study of the population of obscured AGB stars in the MCs to
be made. In this paper we present the results of an attempt to
classify these stars into oxygen and carbon stars, and to study their
luminosity distributions.

Optically bright AGB stars are relatively easy to classify as carbon-
or oxygen-rich from low-resolution spectra in the 500 to 800~nm region.
In the NIR, carbon stars are often distinguished from oxygen stars by
their $(J-K)$ colours (e.g.\ Feast et al.\ 1982). These methods do not
work for obscured AGB stars which are optically too faint, and whose
NIR colours are more dependent on the optical depth of the
circum-stellar envelope (CSE) than on the effective temperature (Feast
1996). For obscured stars, OH maser emission indicates an oxygen-rich
CSE, but few evolved stars in the MCs have detectable OH masers
(Wood et al.\ 1992) and this technique is of limited use in the
LMC. The IRAS $(\left[25\right]-\left[60\right])$ versus
$(\left[12\right]-\left[25\right])$ two-colour diagram may be used
to separate carbon-rich CSEs from oxygen-rich CSEs (van der Veen
1989), but the MCs are too distant for the IRAS instruments to yield
reliable 60~$\mu$m fluxes. Guglielmo et al.\ (1993) demonstrated the
use of combined near- and mid-IR colour-colour diagrams in isolating
carbon stars in the Milky Way. In papers~II and III we confirmed that
the $(K-\left[12\right])$ versus $(H-K)$ or $(J-K)$ diagram can be
used successfully for the MCs. This is the principal method we use
here to chemically classify our MC sample of obscured AGB stars, with
in a few cases an optical spectrum or an OH maser detection too.

In Sect.\ 2 we present $N$-band photometry for 21 obscured AGB stars
in the LMC, 2 in the SMC and the SMC red supergiant (RSG) or
foreground star, VV Tuc. These stars form a subsample of the obscured
AGB candidates in paper~II. In Sect.\ 3 we discuss additional NIR
photometry from SAAO for these sources. Sect.\ 4 describes the
results of a search for OH maser emission from fields in the LMC,
centred at known obscured AGB stars, trying to extend the known sample
of OH/IR stars in the LMC (Wood et al.\ 1992). Sect.\ 5 presents
optical/NIR spectra of obscured AGB stars in the LMC. In Sect.\ 6 we
discuss the $(K-\left[12\right])$ versus $(H-K)$ diagram used to
classify the AGB stars according to the chemical type of their CSEs.
We also study the position of the RSGs in this diagram. In Sect.\ 7
we derive bolometric luminosities, and investigate the time
variability in the $N$-band. The luminosity distributions of the
obscured AGB stars and the relative distributions of the carbon- and
oxygen-rich stars are derived. We discuss the results and summarise
the conclusions.

\section{Mid-infrared imaging photometry}

%
% TABLE 1
%
\begin{table*}[tb]
\caption[]{Names, Heliocentric Julian Dates, $K$ and $N$-band
magnitudes, $(H-K)$ and $(K-\left[12\right])$ colours for the stars in
our $N$-band photometry sample. We adopt $\left[12\right] = N-0.33$
(see text). 1-$\sigma$ error estimates are given.}
\begin{tabular}{lllllllllll}
\hline\hline
IRAS & other & HJD$-$2\,440\,000 & $K$ & $\sigma_K$ & $N$ & $\sigma_N$
& $(H-K)$ & $\sigma_{(H-K)}$ & $(K-\left[12\right])$ &
$\sigma_{(K-\left[12\right])}$ \\
\hline
\multicolumn{11}{l}{\it SMC} \\
00165--7418 & VV Tuc  & 9677.62 &  7.1  & 0.3  &   4.67  & 0.06 & 0.24
& 0.10 &   2.76  & 0.3  \\
00350--7436 &         & 9676.57 &  9.13 & 0.03 &   5.20  & 0.04 & 1.08
& 0.01 &   4.26  & 0.05 \\
01074--7140 & HV12956 & 9676.60 &  9.6  & 0.1  &   4.99  & 0.03 & 0.38
& 0.02 &   4.94  & 0.10 \\
\multicolumn{11}{l}{\it LMC} \\
04286--6937 &         & 9676.63 & 11.07 & 0.10 &   5.99  & 0.07 & 1.76
& 0.02 &   5.41  & 0.12 \\
04374--6831 &         & 9676.67 & 12.42 & 0.05 &   6.09  & 0.07 & 2.46
& 0.02 &   6.66  & 0.09 \\
04407--7000 &         & 9676.70 &  9.17 & 0.05 &   5.30  & 0.05 & 1.11
& 0.02 &   4.20  & 0.07 \\
04496--6958 &         & 9676.72 &  8.84 & 0.02 &   5.03  & 0.04 & 1.40
& 0.01 &   4.14  & 0.04 \\
04498--6842 &         & 9676.73 &  7.51 & 0.01 &   3.70  & 0.03 & 0.53
& 0.01 &   4.14  & 0.03 \\
04539--6821 &         & 9676.76 & 12.7  & 0.1  &   5.81  & 0.06 & 2.7
& 0.2  &   7.22  & 0.12 \\
04557--6753 &         & 9676.78 & 11.56 & 0.02 &   4.94  & 0.03 & 2.12
& 0.07 &   6.95  & 0.04 \\
05003--6712 &         & 9677.71 & 10.8  & 0.2  &   5.89  & 0.10 & 1.56
& 0.10 &   5.24  & 0.22 \\
05009--6616 &         & 9677.68 & 11.00 & 0.05 &   5.23  & 0.02 & 1.75
& 0.02 &   6.10  & 0.05 \\
05099--6740 & TRM023  & 9677.73 & 11.41 & 0.07 & \llap{$>$}6.86 & 0.23
& 1.19 & 0.01 & \llap{$<$}4.88 & 0.24 \\
05112--6755 & TRM004  & 9677.75 & 12.3  & 0.1  &   4.94  & 0.03 & 2.17
& 0.15 &   7.69  & 0.10 \\
05112--6739 & TRM024  & 9677.77 & 12.9  & 0.2  &   5.24  & 0.05 & 2.7
& 0.2  &   7.99  & 0.21 \\
05117--6654 & TRM072  & 9677.79 & 11.39 & 0.03 &   5.45  & 0.13 & 1.70
& 0.10 &   6.27  & 0.13 \\
05128--6455 &         & 9677.80 & 10.36 & 0.02 &   4.93  & 0.04 & 1.53
& 0.02 &   5.76  & 0.04 \\
05190--6748 & TRM020  & 9677.82 & 12.4  & 0.1  &   4.86  & 0.03 & 2.9
& 1.0  &   7.87  & 0.1  \\
05203--6638 & TRM088  & 9677.84 & 10.50 & 0.03 &   6.00  & 0.09 & 1.65
& 0.02 &   4.83  & 0.09 \\
05291--6700 &         & 9677.86 & 10.1  & 0.3  & \llap{$>$}6.99 & 0.28
& 0.97 & 0.10 & \llap{$<$}3.44  & 0.4  \\
05329--6709 & TRM060  & 9676.80 &  9.4  & 0.1  &   3.70  & 0.02 & 2.20
& 0.05 &   6.03  & 0.1 \\
05348--7024 &         & 9676.81 & 14.4  & 1    &   5.17  & 0.02 & 4.9
& 1.0  &   9.56  & 1.0  \\
05360--6648 & TRM077  & 9676.84 & 13.9  & 1    &   6.30  & 0.08 & 2.8
& 0.6  &   7.93  & 1.0  \\
05506--7053 &         & 9676.86 & 12.2  & 0.2  & \llap{$>$}6.14 & 0.11
& 2.4 & 0.2  & \llap{$<$}6.39  & 0.23 \\
\hline
\end{tabular}
\end{table*}

We used the ESO 10~$\mu$m camera TIMMI (K\"{a}ufl et al.\ 1992) at the
3.6m telescope at La Silla on the nights of 1994 November 19/20 and
20/21 to obtain $N$-band photometry ($\lambda_0$ = 10.10$\mu$m,
$\Delta\lambda$ = 5.10$\mu$m). This filter is centred on the silicate
dust feature, which is prominent in oxygen-rich CSEs. We chose a scale
of $0.5^{\prime\prime}$ per pixel, giving a field of view of
$32^{\prime\prime}\times 32^{\prime\prime}$. Because of the very high
background radiation the standard procedure for observing in the
thermal IR is chopping and nodding. We used a chopper throw of
$8^{\prime\prime}$, which insured that the source was in all of the
frames, thereby increasing the signal-to-noise significantly.

Flat-fields were obtained by measuring the flux of a standard star at
13 positions uniformly distributed over the array, and fitting a
two-dimensional parabola to the measured values. This gives reliable
corrections over all but the very edges of the array. We followed the
reduction procedure as it is described in paper~II to derive
magnitudes. This method is based on the sampling of the point-spread
function for each star individually by means of (software-)aperture
photometry with an increasing aperture size. The deduced magnitude
profile is then compared to that of a standard star. In this way we
obtained accurate and reliable magnitude measurements, as well as
reliable error estimates.

The $N$-band magnitudes are listed in Table 1, along with their
1-$\sigma$ error estimates, and the times of mid-exposure. The
exposure times were typically between 10 and 30 min. In the case of a
non-detection in the $N$-band, the 1-$\sigma$ error indicates the
probability that the source is actually brighter than the lower limit
given in Table 1. We refer to Appendix A for a discussion about the
photometric standard stars that are available for the $N$-band.

\section{Near-infrared photometry}

The $N$-band photometry is complemented by near-infrared (NIR)
photometry from a project at SAAO to monitor dust-enshrouded Long
Period Variables (LPVs) in the LMC. At this stage, we only derived
NIR magnitudes and colours for the epoch of $N$-band measurement,
from the monitoring data as it was available in 1996 March.

The NIR magnitudes measured at different epochs were interpolated to
obtain estimates for the 1994 November epoch of the $N$-band
measurement. The errors on the $K$ magnitudes were estimated ``by
eye''. They depend on the degree to which the light curve is sampled,
as well as on the accuracies of the individual measurements. The
$(H-K)$ colour at the time of the $N$-band measurement was estimated
by interpolating the $(H-K)$ colours as they had been measured at the
different epochs, rather than interpolating $H$ and $K$ magnitudes
separately and deriving $(H-K)$ from the interpolated magnitudes. The
two values for the $(H-K)$ colour were always consistent with each
other, indicating a reliable estimation of the accuracies of the
estimated individual $H$ and $K$ magnitudes for the 1994 November
epoch of the $N$-band measurement. The results are presented in Table
1. For the single epoch NIR photometry of IRAS00165--7418 we adopted
formal errors of 0.3 and 0.1 mag on $K$ and $(H-K)$ respectively, in
accordance with the expected variability (see below).

For IRAS05291--6700, IRAS05348--7024, and IRAS05360--6648 the IRAC2
data from paper~II were used. We transformed the magnitudes from the
IRAC2 system to the SAAO system (Carter 1990), using (Lidman 1995):
\begin{equation}
\left( \begin{array}{c} J_{\rm SAAO} \\ H_{\rm SAAO} \\ K_{\rm SAAO}
\end{array} \right) = \left( \begin{array}{rrr} 1.125 & 0 & -0.125 \\
-0.032 & 1 & 0.032 \\ 0.059 & 0 & 0.941 \end{array} \right) \times
\left( \begin{array}{c} J_{\rm IRAC2} \\ H_{\rm IRAC2} \\ K_{\rm
IRAC2} \end{array} \right)
\end{equation}
For the reddest sources, the difference is of the order of a few
tenths of a magnitude. The formal errors include the effects of the
faintness of the source, the difference in epochs between the IRAC2
and TIMMI observations, the expected amplitude of variability, and the
error introduced by the conversion from the IRAC2 to the SAAO
photometric system. For IRAS05348--7024 we assumed an IRAC2 $J$-band
magnitude of $J_{\rm IRAC2} = 24$, to be able to calculate the
transformation to the SAAO system. This value has been chosen to be
consistent with the expected SAAO $J$-band magnitude (see Appendix B).
The error introduced by this assumption will not dominate the error in
the derived NIR magnitudes. The $(H-K)$ colour for IRAS05190--6748 has
been estimated from the $(K-L)$ colour as described in the Appendix,
adopting a formal error of 1.0 magnitude.

\section{A search for OH maser emission from two LMC fields}

The presence of a strong OH maser in a mass-losing star is direct
evidence that the star is oxygen-rich. OH maser emission has previously
been detected from six luminous AGB stars and RSGs in the LMC by Wood et
al.\ (1992). Their detections, with the Parkes 64-m radio telescope with
an OH detection level of $\sim 60$ mJy, suggested that the expansion
velocities of OH/IR stars in the LMC are substantially lower than for
OH/IR stars in the Milky Way. Zijlstra et al.\ (1996) and van Loon et
al.\ (1996), however, found a difference of only $\sim20$ to 30\% at
most. The OH (non-)detections demonstrated that the LMC OH/IR stars are
likely to have optically thinner CSEs than their galactic counterparts.
Less IR pumping, together with a lower oxygen abundance, may then lead
to the low OH maser flux densities of the LMC sources.

%
% FIGURE 1
%
\begin{figure*}[tb]
\centerline{\psfig{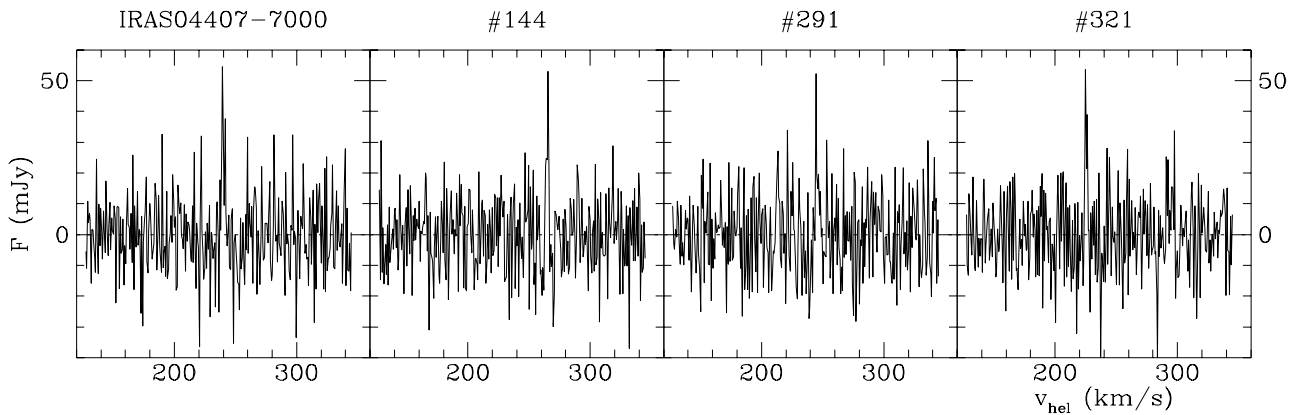}}
\caption[]{Spectra of the 1612 MHz OH maser detection of
IRAS04407--7000 (leftmost), and the candidate detections {\#}144
(center left), {\#}291 (center right), and {\#}321 (rightmost)}
\end{figure*}

In order to test whether OH maser emission could be detected from
sources with lower infrared luminosities, we decided to search for OH
maser emission at 1612 MHz, in two LMC fields of size $0.5 \times 0.5$
square degrees. The selected fields were centred at the two IRAS
sources IRAS04407--7000 and IRAS05112--6755. IRAS04407--7000 is one of
the brightest AGB stars in the LMC with an IRAS 12~$\mu$m flux density
of 0.81 Jy, while IRAS05112--6755 is also an AGB star (paper~II), with
an IRAS 12~$\mu$m flux density of 0.41 Jy.

\subsection{OH Observations}

The OH maser observations were made with the Australia Telescope
Compact Array (ATCA) on 1994 September 5 to 7. The ATCA is an
east-west array of six 22-m antennas which has a maximum baseline of
6~km. For a wavelength of 18~cm, the angular resolution is
approximately 6 arcsec.

The two fields, centred on the positions of IRAS05112--6755 and
IRAS04407--7000, were observed for total on-source integration times
of 10 and 3 hrs, respectively. The total spectral bandpass of 4~MHz
was centred at 1611 MHz. It was split into 1024 spectral channels,
giving a channel separation of 0.71 km~s$^{-1}$, and a velocity
resolution of 0.85 km~s$^{-1}$. The data were corrected for
atmospheric amplitude and phase variations using observations of
strong nearby continuum sources. To calibrate the flux density scale,
the primary calibrator source, 1934--638 was also observed. This was
taken to have a flux density of 14.8 Jy at 18~cm.

The data were reduced using routines in the AIPS and MIRIAD
radio-astronomy packages. As considerable interference from the
Russian Glonass satellites was present during the observations, the
visibility data for the shorter baselines were first edited to remove
sections containing strong interference. Glonass interference signals
near 1612 MHz are generally detected on baselines below one km. After
removing the interference the data were then Fourier transformed to
the image plane.

\subsection{Search strategy}

To search for detections we constructed data cubes on the inner
$16.3^{\prime}\times 16.3^{\prime}$ of the fields centred on each
of the two IRAS sources, covering a velocity range of 170 to 340
km~s$^{-1}$. Each cube consisted of $512 \times 512 \times 230$ pixels
of size $2.5^{\prime\prime}\times 2.5^{\prime\prime}\times
0.73$~km~s$^{-1}$. Natural weighting was applied in deriving
resolution matched spectra from this cube. In this way we achieved
final RMS noise levels (1-$\sigma$) of 12~mJy and 6~mJy for the fields
on IRAS04407--7000 and IRAS05112--6755 respectively.

The cubes were searched for all spikes above 4.7~$\sigma$. This
resulted in about 250 candidates which were inspected by eye. The
number is consistent with Gaussian distributed noise. All of the
spikes were single velocity-channel spikes and disregarded as noise.
A filter was applied to the data allowing us to lower the threshold
to 3.5~$\sigma$. This filter demanded that three neighbouring velocity
channels had a combination of fluxes equal or exceeding 1, 0.3 and 0.2
times the 3.5~$\sigma$ threshold. Around 450 candidates per field were
recovered. This number is again consistent with random noise.
Inspection of all individual spectra resulted in a list of a few dozen
candidates, the best of which is {\#}321 (our nomenclature).

\subsection{OH detections}

%
% TABLE 2
%
\begin{table}[tb]
\caption[]{Names, radio positions, and heliocentric peak velocities
(in km~s$^{-1}$) of the (possible) OH detections.}
\begin{tabular}{llll}
\hline\hline
Name & RA (2000.0) & Dec (2000.0) & $v_{\rm hel}$ \\
\hline
IRAS04407--7000 & $04^{\rm h}40^{\rm m}28.5^{\rm s}$ &
$-69^{\circ}55^{\prime}14^{\prime\prime}$ & 239 \\
{\#}144         & $04^{\rm h}40^{\rm m}19.3^{\rm s}$ &
$-69^{\circ}57^{\prime}44^{\prime\prime}$ & 266 \\
{\#}291         & $04^{\rm h}40^{\rm m}34.3^{\rm s}$ &
$-69^{\circ}51^{\prime}19^{\prime\prime}$ & 244 \\
{\#}321         & $04^{\rm h}40^{\rm m}14.0^{\rm s}$ &
$-69^{\circ}50^{\prime}31^{\prime\prime}$ & 226 \\
\hline
\end{tabular}
\end{table}

The candidates were checked for IRAS counterparts. None were found,
except for the known obscured AGB star IRAS04407--7000 (see paper~II).
The positions of two other candidates are coincident with stars
visible on ESO-POSS plates: {\#}144 and {\#}291. The spectra of four
candidate OH-maser sources are shown in Fig.\ 1, and their parameters
are given in Table 2. All detections peak with a flux density of about
0.05 Jy (a bit over 4~$\sigma$). We conclude that one new OH/IR star
(IRAS04407--7000) was detected, and that there may be up to a few
dozen more OH/IR stars at a flux level of $\sim$40 to 50 mJy. Current
technology is able to detect only the brightest stellar masers at the
distance of the MCs (Wood et al.\ 1992; van Loon et al.\ 1996). Any
significant improvement in sensitivity would therefore be expected to
drastically increase the number of stellar masers known in the MCs.

\section{Optical/NIR spectroscopy of IR stars in the LMC}

%
% FIGURE 2
%
\begin{figure*}[tb]
\centerline{\psfig{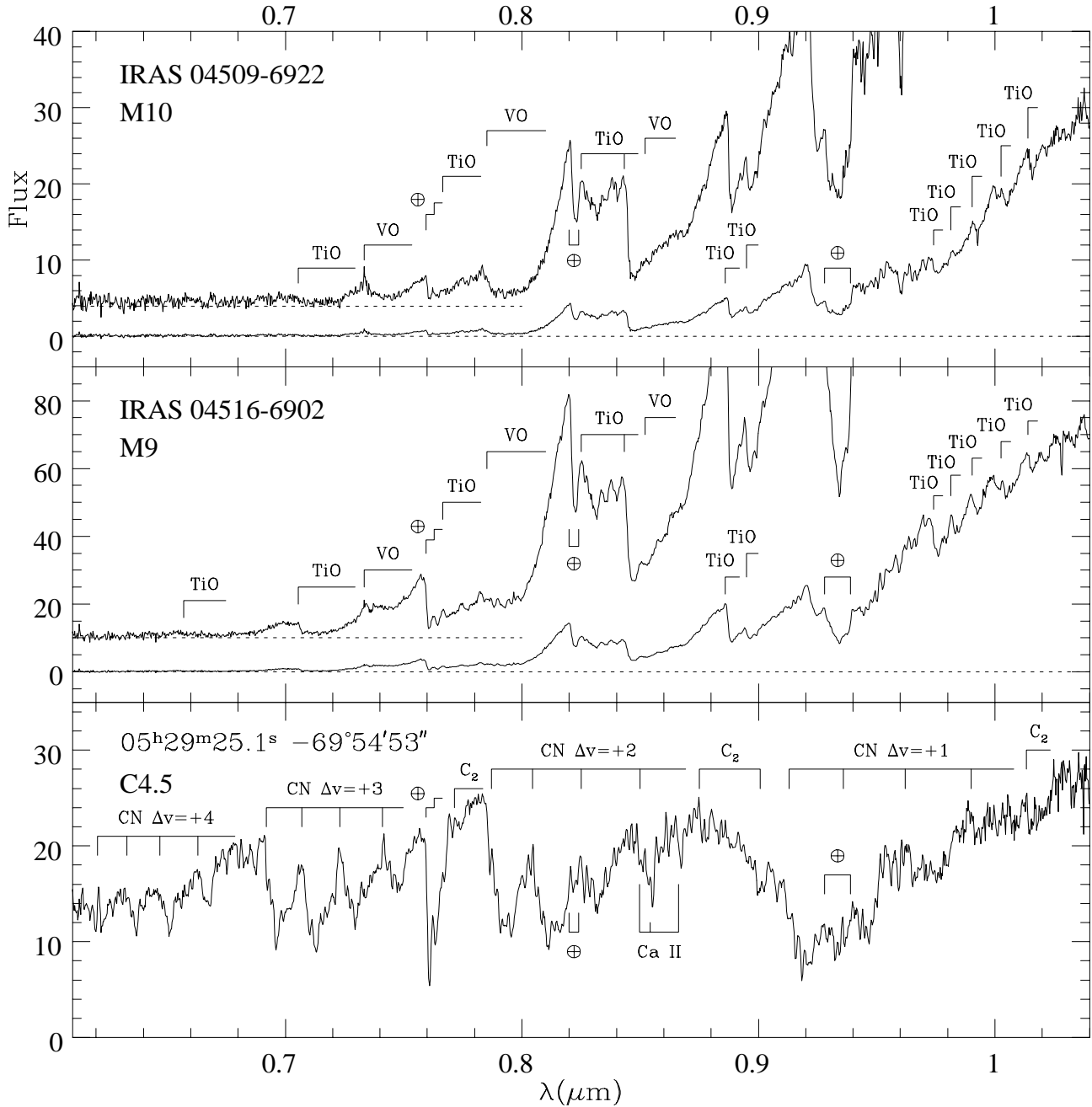}}
\caption[]{Spectra of the IR stars IRAS04509--6922 and
IRAS04516--6902, and of a serendipitously discovered carbon star of
which the position is given in 2000 coordinates accurate to within
$\sim2^{\prime\prime}$. The spectra of the IR stars are displayed a
second time, multiplied by 5 (and with a small offset in flux) to show
better the heavily extinguished part around 0.7~$\mu$m. The flux scales
have not been calibrated. The main spectral features are indicated,
including the strongest telluric absorption features}
\end{figure*}

We used the ESO 3.5m New Technology Telescope at La Silla on the night
of 1996 February 4 to obtain low resolution (R=500) spectra, between
0.6 and 1~$\mu$m, of the stars IRAS04509--6922 and IRAS04516--6902, to
chemically classify their photospheres. The stars were identified from
their very red colours on $V$- and $I$-band acquisition images. The
CCD frames were corrected for the electronic offset (bias) and for the
relative pixel response (flatfield). The sky-subtracted spectra were
then corrected for the wavelength dependence of the instrumental
response, and for atmospheric extinction.

\subsection{IRAS04509--6922 and IRAS04516--6902}

The spectra of IRAS04509--6922 and IRAS04516--6902 are presented in
Fig.\ 2. The strong TiO bands leave no doubt about their photospheres
being oxygen rich. We show below that they are severely obscured by
dusty CSEs. They are both luminous, large amplitude variables, with
periods of 1290 and 1090 days respectively (Wood et al.\ 1992), making
them good OH maser candidates. IRAS04516--6902 has not been searched
for OH maser emission, while for IRAS04509--6922 Wood et al.\ (1992)
derive an upper limit of 0.04~Jy, which is comparable to the fluxes
from the new detection(s) discussed above. The only extra-galactic
OH/IR star of which an optical spectrum had been taken so far is the
extremely bright red supergiant IRAS04553--6825 (Elias et al.\ 1986).

We classify the two IR stars to half a subclass accuracy on the basis
of the relative strengths of the various molecular bands at about
0.71, 0.77, 0.83, 0.84, and 0.89~$\mu$m (TiO) and at about 0.74, 0.79,
and 0.86~$\mu$m (VO), comparing with the representative spectra in
Turnshek et al.\ (1985) and especially Fluks et al.\ (1994). We find
very late types of M10 (IRAS04509--6922) and M9 (IRAS04516--6902). The
relative strengths of the TiO bands in the 0.97 to 1.02~$\mu$m region,
and the absence of FeH absorption at 0.99~$\mu$m indicate that these
stars are giants not galactic dwarfs (Couture \& Hardy 1993). Fluks et
al.\ (1994) assign effective temperatures of $\sim 2700$~K and
$2500$~K to spectral types of M9 and M10 respectively, but it is not
known how this depends on metallicity. Feast (1996) suggests that Mira
variables may be different from other M-type stars in their relation
between spectral type and effective temperature.

To quantify the reddening by the CSE, we compare the intensity levels
in the spectrum at 0.70~$\mu$m and 0.84~$\mu$m between the programme
star and a comparison star of the same spectral type. The spectral
slope at these pseudo-continuum wavelengths is relatively flat over
some 0.01~$\mu$m, and the result is therefore not very sensitive to
the spectral resolution. There may be a slight metallicity effect,
introduced by the TiO absorption bands at 0.84~$\mu$m, but this is
probably small compared to the effects of extinction. Differential
slit losses are not important, as the slit was positioned at the
$I$-band stellar image and the spectra were taken at air masses smaller
than 1.5.

%
% TABLE 3
%
\begin{table}[tb]
\caption[]{Ratio of intensity levels in the spectrum at 0.70~$\mu$m
and 0.84~$\mu$m. They are estimated from the intrinsic spectra from
Fluks et al.\ (1994), and from the spectra of the oxygen-rich IR stars
shown here. For IRAS04509--6922 and IRAS04516--6902 the derived
extinction in magnitudes for a Johnson $V$ measurement is given.}
\begin{tabular}{llll}
\hline\hline
Star            & Spec.Type & $I_{0.70}/I_{0.84}$ & $A_V$ \\
\hline
Fluks et al.\   &    M8     &        0.340          &     0       \\
Fluks et al.\   &    M9     &        0.250          &     0       \\
Fluks et al.\   &    M10    &        0.174          &     0       \\
IRAS04509--6922 &    M10    &        0.09           &     3.4     \\
IRAS04516--6902 &    M9     &        0.10           &     4.7     \\
\hline
\end{tabular}
\end{table}

We adopt an extinction curve $A_\lambda/E_{(B-V)}$ (Fluks et al.\ 1994,
and references therein), and correct the $A_V = R \times E_{(B-V)}$ for
the dependence on spectral type (Fluks et al.\ 1997, who calibrate to
$R = 3.1$ for an O7.5V star). We measured the ratio of the intensities,
at 0.70~$\mu$m and 0.84~$\mu$m, for the comparison spectra from Fluks
et al.\ (1994) and for the programme stars: IRAS04509--6922 and
IRAS04516--6902. From these ratios the extinction, $A_V$, is derived in
magnitudes as it would be measured in the Johnson $V$-band (Table 3).
The $A_V$ estimation is accurate to within a magnitude. Thus it is
clear that both stars suffer from significant circumstellar extinction,
although less than do typical galactic OH/IR stars which have $A_V \gg
10$ mag (Habing 1996).

\subsection{A cluster carbon star near IRAS05298--6957}

The OH/IR star IRAS05298--6957 (Wood et al.\ 1992) could not be
identified in the $V$- and $I$-band snapshots. It was probably
considerably fainter than $I=20$ mag (60 s acquisition exposure). Its
location in the small star cluster HS327 (Hodge \& Sexton 1966)
results in severe crowding and contributes to the difficulty in
finding it.

However, a very red (in $V-I$) star was found at the Northern rim of
the same cluster, about $22^{\prime\prime}$ from the position of the
OH/IR. A spectrum of this star was obtained, in the same way as
described in section 5.0, and is illustrated in Fig.\ 2. The numerous
CN bands, e.g.\ at about 0.63, 0.65, 0.67, 0.70, 0.71, 0.73, 0.75,
0.79, and 0.92~$\mu$m, and the weak C$_2$ absorption at about
1.02~$\mu$, leave no doubt that it is a carbon star. The very red
colours which resulted in its discovery are probably a consequence of
very strong C$_2$ bands shortward of 0.563~$\mu$m. This absorption
will considerably decrease the $V$-band flux.

%
% FIGURE 3
%
\begin{figure*}[tb]
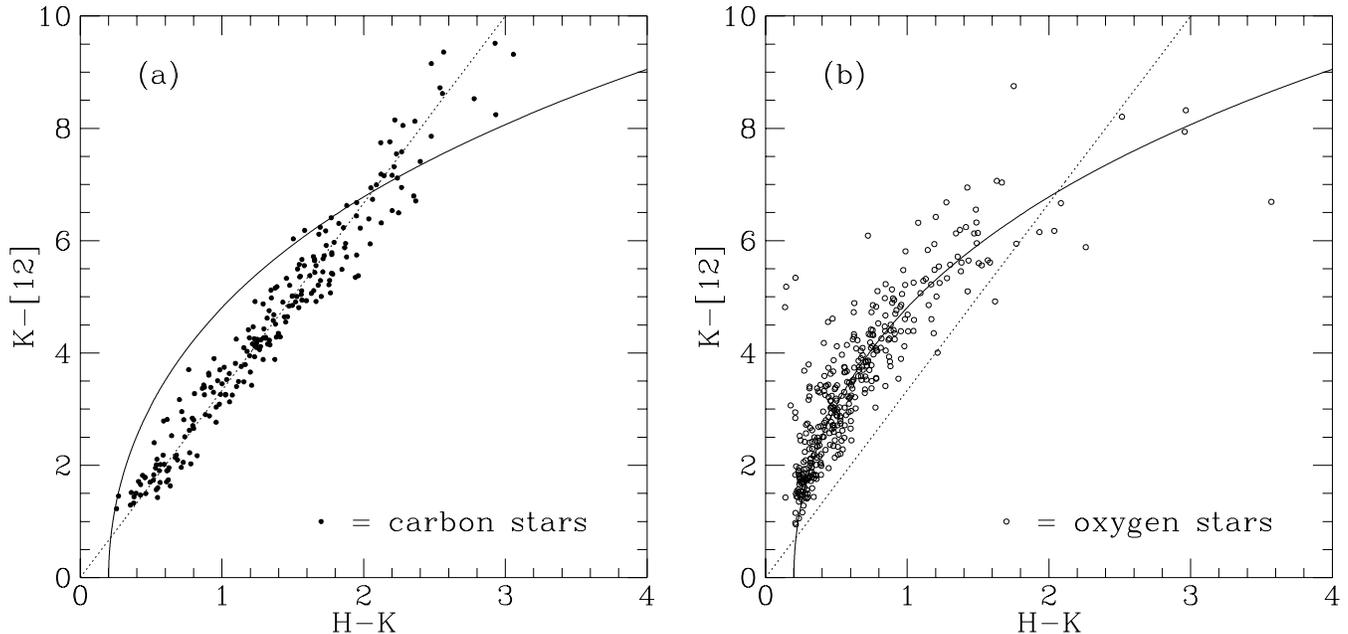

\centerline{\hbox{
\psfig{figure=3065.f3a,width=90mm}
\psfig{figure=3065.f3b,width=90mm}}}
\caption[]{$(K-\left[12\right])$ versus $(H-K)$ diagram for the carbon
stars (a) and oxygen stars (b) of the galactic sample of Guglielmo et
al.\ (1993). The dotted line and the solid curve are the empirical
carbon- and oxygen-star sequences, respectively (see text)}
\end{figure*}

Unfortunately, our spectral coverage does not include the 0.44 to
0.6~$\mu$m region, upon which most spectral classification is based
(e.g.\ Keenan 1993). The star is not a galactic dwarf (cf.\ Green
1997): the CaH bands at about 0.638 and 0.639~$\mu$m are absent, the
K~{\sc i} lines at 0.767 and 0.770~$\mu$m are weak, and the FeH line
at 0.99~$\mu$m is weak or absent, all of which are enhanced in late-M
dwarfs (Jaschek \& Jaschek 1987; Turnshek et al.\ 1985; Couture \&
Hardy 1993). The equivalent width of the Ca~{\sc ii} triplet
component at 0.866~$\mu$m is $W_\lambda \sim 4.7\pm0.5$~\AA,
suggesting it is a giant not a dwarf (Danks \& Dennefeld 1994 and
references therein). This is confirmed by the absence of strong
absorption by the Na~{\sc i} doublet around 0.819~$\mu$m (Alloin \&
Bica 1989). A $^{13}$C isotope super-enhancement, characteristic of
J-type carbon stars, is not observed: the CN and C$_2$ bands at 0.606
and 0.619~$\mu$m are much stronger than their $^{13}$C isotopic
equivalents at 0.626 and 0.617~$\mu$m (Richer et al.\ 1979). H$\alpha$
is only barely visible, unlike the strong H$\alpha$ absorption seen in
CH stars (Barnbaum et al.\ 1996). The Ba~{\sc ii} absorption at
0.65~$\mu$m is much weaker than in barium stars. The strong CN
absorption between $\sim$0.6 and 0.8~$\mu$m is suggestive of an N-type
star (i.e.\ a star on the thermally pulsing AGB) rather than of an
R-type. We conclude that the spectral type is C4.5, estimated to
half a subtype accuracy by comparison with spectra from Turnshek et
al.\ (1985) and from Barnbaum et al.\ (1996).

The observed wavelengths of the Ca~{\sc ii} triplet lines suggest a
large radial velocity for the star ($\sim 300 \pm 100$ km~s$^{-1}$)
consistent with membership of the LMC. The flux scales of the spectra
are not calibrated absolutely, mainly because of unknown slit losses
and variable seeing. With uncalibrated acquisition images and the
lack of NIR photometry, the bolometric luminosity of the carbon star
remains unknown. The apparent association of both an OH/IR star and a
carbon-star with a cluster is sufficiently interesting that it is
important to ascertain if both are actually members. If they are, then
the implications of a single population producing an AGB carbon-star
and an OH/IR star at the same time will need serious consideration.

\section{Carbon stars in a $(K-\left[12\right])$ versus $(H-K)$ diagram}

\subsection{The Milky Way}

In Fig.\ 3 we present the $(K-\left[12\right])$ versus $(H-K)$
diagnostic diagram. This is used to distinguish between carbon- and
oxygen-rich stars within the galactic sample of carbon stars (Fig.\ 3a)
and oxygen stars (Fig.\ 3b) that Guglielmo et al.\ (1993) selected to
test the diagnostic value of IR colour-colour diagrams. We adopt
$\left[12\right] = -2.5 \log(S_{12}/28.3)$, where $S_{12}$ is the flux
density in Jy in the IRAS 12~$\mu$m band (IRAS Explanatory Supplement
1988). The original NIR photometry of Guglielmo et al.\ is on the ESO
system, but we transformed their data to the SAAO system (Carter 1990).
The carbon-star sequence (dotted straight line) is described by the
empirical relation
\begin{equation}
(H-K) = 0.3 \times (K-\left[12\right])
\end{equation}
It is remarkable how well this extremely simple relation holds. The
oxygen star sequence (solid curved line) approximately satisfies
\begin{equation}
(H-K) = 0.2 + 0.03 \times (K-\left[12\right])^2 + 0.0002 \times
(K-\left[12\right])^4
\end{equation}
Although less simple than for the carbon-star sequence, the relation
for oxygen stars is also described by a smooth polynomial. The
deviation of the oxygen stars from the carbon star sequence for 
$(K-\left[12\right]) < 7$ mag is caused by the additional emission of
the 10~$\mu$m SiO feature from the dusty CSE, while the deviation for
$(K-\left[12\right]) > 7$ mag is due to additional absorption by the
10~$\mu$m SiO feature.

The data of Guglielmo et al.\ (1993) show that in the region of the
$(K-\left[12\right])$ versus $(H-K)$ diagram at $5 < (K-\left[12\right])
< 9$ mag and $1.3 < (H-K) < 3$ mag there are many more carbon stars (86)
than oxygen stars (30). In Fig.\ 3 carbon stars are found rather evenly
spread along their sequence, whereas oxygen stars are rarely found with
$(H-K)>2$ mag. Moreover the scatter of the most obscured oxygen stars in
the $(K-\left[12\right])$ versus $(H-K)$ diagram is substantial.

\subsection{Magellanic Clouds}

In paper~II we showed that the $(K-\left[12\right])$ versus $(H-K)$
diagnostic diagram may also be applicable to our sample of obscured
AGB stars in the LMC. However, the uncertainties in the IRAS fluxes
were too large to enable an individual separation into carbon and
oxygen stars. Now that we have obtained accurate $N$-band magnitudes
we are able to perform a diagnosis of the C/O abundance ratio for
individual stars in our sample. Comparison of the $N$-band magnitudes
with the IRAS 12~$\mu$m flux densities (paper~II) results in
$(N-\left[12\right]) = 0.33 \pm 0.11$ mag, with no evidence for a
colour term. The uncertainty given is the uncertainty in the
estimation of the mean. Individual values can be off by a magnitude,
but this we attribute to variability of the sources between the IRAS
and $N$-band epochs (see section on variability) rather than
uncertainties in the derived relation. The derived
$(K-\left[12\right])$ colours are listed in Table 1 for the epochs of
the $N$-band measurements.

%
% FIGURE 4
%
\begin{figure}[tb]
\centerline{\psfig{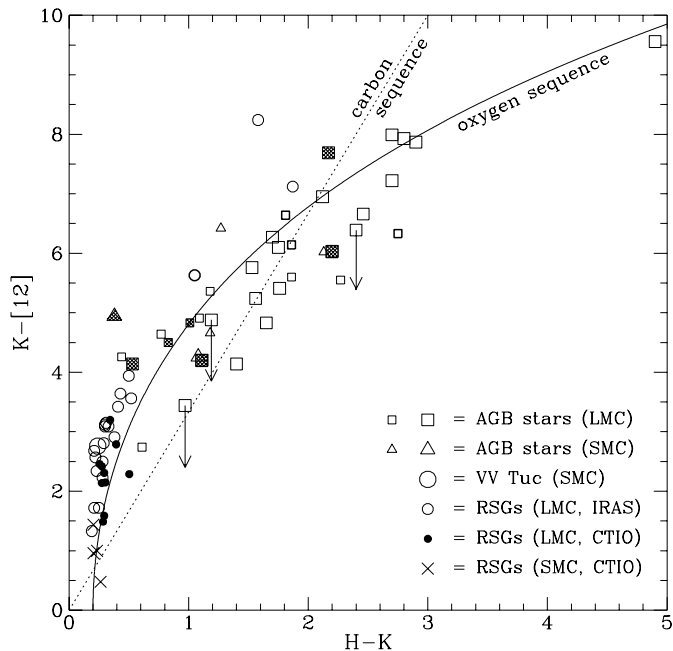}}
\caption[]{$(K-\left[12\right])$ versus $(H-K)$ diagram for the stars of
our sample (large symbols), and the other stars from paper~II (small
symbols). All NIR data is on the SAAO photometric system. We adopt
$\left[12\right] = -2.5 \log(S_{12}/28.3)$, where $S_{12}$ is the flux
density in Jy in the IRAS 12~$\mu$m band (IRAS Explanatory Supplement
1988). Highlighted are OH masers (bold-faced symbols), and stars that
are classified as oxygen stars either from the narrow-band photometry of
paper~II or from our optical/NIR spectroscopy (shaded symbols). Also
plotted are RSGs in the LMC (solid dots) and SMC (crosses) with
10~$\mu$m photometry obtained at CTIO (Elias et al.\ 1985). The dotted
line and the solid curve are the galactic carbon- and oxygen-star
sequences respectively}
\end{figure}

All the stars in our sample are displayed in the $(K-\left[12\right])$
versus $(H-K)$ diagram (Fig.\ 4, large symbols), superimposed with the
galactic carbon-star sequence (dotted straight line) and the galactic
oxygen-star sequence (solid curved line). We also display the other
AGB stars and the RSGs from the LMC and SMC as listed in paper~II (Fig.\
4, small symbols). Elias et al.\ (1985) presented 10~$\mu$m photometry
of a small subset of their sample of RSGs in the MCs, obtained at CTIO.
We treated their 10~$\mu$m magnitudes as if they were TIMMI $N$-band
magnitudes, and they are included in Fig.\ 4 for comparison (solid dots
and crosses for LMC and SMC members, respectively). All data are on the
SAAO photometric system, after applying transformation equations from
Carter (1990) and McGregor (1994) where required. Some stars are
confirmed to be oxygen rich, by our narrow-band photometry (paper~II) or
by our optical/NIR spectroscopy (shaded symbols), or by the detection of
OH-maser radiation (bold symbols).

\subsection{Classification by chemical type}

%
% TABLE 4
%
\begin{table}[tb]
\caption[]{Classification of the stars in our $N$-band photometry sample
as oxygen rich (O), carbon rich (C) or unknown (OC).}
\begin{tabular}{lll}
\hline\hline
 O          & OC          & C           \\
\hline
\multicolumn{3}{l}{\it Small Magellanic Cloud} \\
00165--7418 &             &             \\
            & 00350--7436 &             \\
01074--7140 &             &             \\
\multicolumn{3}{l}{\it Large Magellanic Cloud} \\
            & 04286--6937 &             \\
04374--6831 &             &             \\
04407--7000 &             &             \\
            &             & 04496--6958 \\
04498--6842 &             &             \\
04539--6821 &             &             \\
            & 04557--6753 &             \\
            & 05003--6712 &             \\
            & 05009--6616 &             \\
            & 05099--6740 &             \\
05112--6755 &             &             \\
            & 05112--6739 &             \\
            & 05117--6654 &             \\
            & 05128--6455 &             \\
05190--6748 &             &             \\
            &             & 05203--6638 \\
            &             & 05291--6700 \\
05329--6709 &             &             \\
05348--7024 &             &             \\
05360--6648 &             &             \\
05506--7053 &             &             \\
\hline
\end{tabular}
\end{table}

We tentatively classify the stars of our $N$-band photometry sample
as oxygen rich or carbon rich. The results are given in Table 4.
Three stars are probably carbon rich (C). Nine other stars have been
labeled ``OC'' because the type could not be determined. They are
situated in a region of the $(K-\left[12\right])$ versus $(H-K)$
diagram where both carbon and oxygen stars are found --- especially
in the region around the carbon-star sequence between $(H-K) = 1.3$
and $(K-\left[12\right]) = 5$ mag, and $(H-K) = 3$ and
$(K-\left[12\right]) = 9$ mag. All other stars of our sample are
probably oxygen rich (O).

We sample well up to the intersection of the oxygen- and carbon-star
sequences at $(K-\left[12\right]) \sim 7$ mag, where the 10~$\mu$m SiO
feature in oxygen stars is thought to go into absorption.
Thus, at $(H-K) \sim$ 5 mag IRAS05348--7024 must be an oxygen star
with strong 10~$\mu$m absorption. The apparent scarcity of stars with
$(H-K) > 3$ mag is an effect of incompleteness, as the deepest NIR
search so far (paper~III) has revealed several of these highly
obscured stars.

OH maser emission was detected from IRAS05329--6709 (Wood et al.\
1992) and from IRAS04407--7000 (see above). This establishes their
status as oxygen rich, since strong OH maser emission is not expected
from carbon stars. The moderate expansion velocity, of about 12
km~s$^{-1}$, of the OH-masing shell around IRAS05329--6709 suggests
that it is an AGB star, not a red supergiant (RSG). The narrow-band
mid-IR photometry (paper~II) for these two stars, as well as for
IRAS01074--7140, IRAS04498--6842 and IRAS05112--6755 also lead us to
classify these stars as oxygen rich. A mid-IR spectrum of
IRAS05329--6709 (Groenewegen et al.\ 1995) directly confirms it to be
oxygen rich. Convolution of this spectrum with the response curves of
the TIMMI narrow-band filters, and comparison with a flat spectrum
representing an A0~V star yields colours
$(\left[9.8\right]-\left[11.3\right]) = 0.33$ mag and
$(\left[N\right]-\left[9.8\right]) = -0.22$ mag with a crudely
estimated accuracy of 0.1 mag. This agrees within the estimated errors
with the colours, $(\left[9.8\right]-\left[11.3\right]) = 0.41$ mag
and $(\left[N\right]-\left[9.8\right]) = -0.02$ mag, derived from the
narrow-band photometry (paper~II).

The two stars IRAS04509--6922 and IRAS04516--6902 were shown, from the
spectra discussed above, to be oxygen rich. Their $(H-K)$ and
$(K-\left[12\right])$ colours from Wood et al.\ (1992), after
transformation to the SAAO system (McGregor 1994), place them very
nicely on the oxygen-rich sequence, in a region where the 10~$\mu$m
SiO feature is in emission. These stars are thus indeed experiencing
mass loss, confirming that their extinction is circumstellar rather
than interstellar.

The position of the SMC source IRAS00165--7418 (VV Tuc) at
$(K-\left[12\right]) = 2.76$ and $(H-K) = 0.24$ is consistent with its
classification as a mass-losing RSG in the SMC (see next section),
rather than as a foreground star. The position of the SMC source
IRAS01074--7140 in the $(K-\left[12\right])$ versus $(H-K)$ diagram
suggests it may be a post-AGB star (cf.\ van Loon et al.\ 1997). The
blue $(H-K)$ colour may reflect a higher effective temperature than an
AGB star has, while there is still a considerable amount of 12~$\mu$m
emission from the CSE. Alternatively, it may still be on the AGB, but
have interrupted its mass loss recently following a thermal pulse. A
similar scenario has been suggested for certain galactic stars by
Whitelock et al.\ (1995).

The $(K-\left[12\right])$ versus $(H-K)$ diagram indicates that
IRAS04496--6958 is a very good candidate carbon star. A preliminary
analysis of a 3~$\mu$m spectrum that we obtained with the IR
spectrometer at the CTIO 4m telescope in December 1996 indeed
confirms the carbon-rich nature of the CSE of IRAS04496--6958.

We take the general agreement between the $(K-\left[12\right])$ versus
$(H-K)$ diagram and more direct observational diagnostics as a
justification for the use of the $(K-\left[12\right])$ versus $(H-K)$
diagram in chemically classifying obscured AGB stars in the MCs.

\subsection{Red supergiants}

The RSGs in the LMC, detected by IRAS, form a sequence that is
slightly shifted to bluer $(H-K)$ or redder $(K-\left[12\right])$ with
respect to the galactic sample of the stars from Guglielmo et al.\
(1993). The sample of Guglielmo et al.\ contains mainly AGB stars with
few RSGs, because their stars were selected outside of the galactic
plane. Thus mass-losing RSGs appear to have different IR colours than
mass-losing AGB stars. At a given mass-loss rate, RSGs may have
similar 12~$\mu$m flux densities to AGB stars, but the RSGs may be
less obscured because of a greater inner radius of their CSEs. This
would yield similar $(K-\left[12\right])$ colours, but smaller $(H-K)$
colours for the RSGs. This may also explain the large IR excess of the
RSG IRAS05216--6753, compared to its NIR extinction: $(H-K) = 1.58$
and $(K-\left[12\right]) = 8.24$ mag. The effect is less visible in
the IR colours of the RSGs from Elias et al.\ (1985), because these
stars have lower mass-loss rates and therefore less 10~$\mu$m emission
from a CSE.

Some stars we have classified as being on the AGB may actually be
RSGs, and vice versa. The candidate AGB star IRAS04498--6842 may be an
example of such a misclassification, as it was very luminous at the
epoch of the $N$-band measurement (see below): it has
$(K-\left[12\right])$ and $(H-K)$ colours which perfectly match the
corresponding colours of the RSGs in the LMC. The same is true for
IRAS05316--6604 (WOH SG374) at $(H-K) = 0.44$ and $(K-\left[12\right])
= 4.26$ mag.

The datum at $(H-K) = 0.61$ and $(K-\left[12\right]) = 2.74$ mag is
WOH SG061 (SHV0453582--690242). Westerlund et al.\ (1981) selected
this star on the basis of objective-prism spectra indicating an M-type.
This spectral type was confirmed by Hughes \& Wood (1990) using
low-dispersion optical/NIR spectroscopy. In paper~II we found it to
be an AGB star, rather than a RSG. In the $(K-\left[12\right])$ versus
$(H-K)$ diagram WOH SG061 follows the sequence for oxygen stars on the
AGB as observed in the Milky Way. As the red $(H-K)$ colour of WOH
SG061 with respect to the RSGs in the LMC is not due to a carbon-rich 
CSE or photosphere, we conclude that indeed mass-losing AGB stars in
the LMC have redder $(H-K)$ colours than mass-losing RSGs in the LMC.

\section{Luminosities}

\subsection{Variability}

%
% TABLE 5
%
\begin{table*}[tb]
\caption[]{Difference of the near and mid-IR magnitudes and colours
between the 1994 November and 1993 December measurements for the stars
in our sample for which we have $N$-band measurements at both epochs.
$\Delta$ refers to (1994 value)--(1993 value). We also derived an
estimation for the typical magnitude difference between minimum and
maximum (amplitude, see text).}
\begin{tabular}{lrlllllll}
\hline\hline
IRAS & $K$ ('93) & $\Delta K$ & $(H-K)$ ('93) & $\Delta (H-K)$ & $N$
('93) & $\Delta N$ & $(K-\left[12\right])$ ('93) & $\Delta
(K-\left[12\right])$ \\
01074--7140 & $9.90 \pm 0.10$ & \llap{$-$}$0.30 \pm 0.14$ & $0.44 \pm
0.05$ & \llap{$-$}$0.06 \pm 0.05$ & $5.78 \pm 0.04$ & \llap{$-$}$0.79
\pm 0.05$ & $4.45 \pm 0.11$ & $0.49 \pm 0.15$ \\
04407--7000 & $8.93 \pm 0.01$ & $0.24 \pm 0.05$ & $0.89 \pm 0.01$ &
$0.22 \pm 0.02$ & $4.76 \pm 0.02$ & $0.54 \pm 0.05$ & $4.50 \pm 0.03$
& \llap{$-$}$0.30 \pm 0.07$ \\
04498--6842 & $7.84 \pm 0.01$ & \llap{$-$}$0.33 \pm 0.01$ & $0.64 \pm
0.01$ & \llap{$-$}$0.11 \pm 0.01$ & $3.67 \pm 0.02$ & $0.03 \pm 0.04$
& $4.50 \pm 0.03$ & \llap{$-$}$0.36 \pm 0.04$ \\
05112--6755 & $13.20 \pm 0.10$ & \llap{$-$}$0.90 \pm 0.14$ & $1.93 \pm
0.15$ & $0.24 \pm 0.22$ & $5.56 \pm 0.05$ & \llap{$-$}$0.62 \pm 0.06$
& $7.97 \pm 0.11$ & \llap{$-$}$0.28 \pm 0.15$ \\
05329--6709 & $9.80 \pm 0.4$ & \llap{$-$}$0.04 \pm 0.4$ & $2.24 \pm
0.20$ & \llap{$-$}$0.04 \pm 0.20$ & $3.91 \pm 0.02$ & \llap{$-$}$0.21
\pm 0.03$ & $6.22 \pm 0.4$ & \llap{$-$}$0.19 \pm 0.4$ \\
\hline
amplitude & & $1.1$ & & $0.4$ & & $1.4$ & & $1.0$ \\
\hline
\end{tabular}
\end{table*}

%
% TABLE 6
%
\begin{table*}[tb]
\caption[]{$J$, $H$, $K$, and $L$-band magnitudes, and 12 and
25~$\mu$m flux densities for the epoch of the $N$-band measurements,
the magnitude difference between the 12~$\mu$m flux densities at the
$N$-band epoch and the 12~$\mu$m flux densities from IRAS, and the
derived apparent and absolute bolometric magnitudes and luminosities
for the stars for which we present $N$-band photometry. Brackets
denote values derived from empirical colour relationships (see
Appendix).}
\begin{tabular}{lllllllclll}
\hline\hline
IRAS & $J$ & $H$ & $K$ & $L$ & $S_{12}$ (Jy) & $S_{25}$ (Jy) &
$(\left[12\right]_N - \left[12\right]_{\rm IRAS})$ & $m_{\rm bol}$ &
$M_{\rm bol}$ & $\log(L/L_{\sun}$) \\
\hline
00165--7418 &   8.30  &   7.40  &  7.10 &   6.80  & 0.52   &  0.33
& 0.00  & 9.66 & $-9.12$ & 5.54 \\
00350--7436 &  11.50  &  10.21  &  9.13 &   7.79  & 0.32   &  0.23
& \llap{$-$}0.07  & 11.96 & $-6.82$ & 4.62 \\
01074--7140 &  11.00  &  10.10  &  9.70 &   9.00  & 0.39   &  0.42
& 0.06  & 11.83 & $-6.95$ & 4.67 \\
 & & & & & & & & & & \\
04286--6937 &  15.30  &  12.82  & 11.07 &   9.00  & 0.15   &  0.12
&   0.39  & 13.18 & $-5.29$ & 4.00 \\
04374--6831 & (18.04) &  14.87  & 12.42 &   9.59  & 0.14   &  0.13
&   0.33  & 13.48 & $-4.99$ & 3.88 \\
04407--7000 &  11.93  &  10.26  &  9.17 &   7.93  & 0.29   &  0.24
&   1.11  & 12.07 & $-6.40$ & 4.45 \\
04496--6958 &  12.12  &  10.23  &  8.84 &   7.31  & 0.37   &  0.33
& 0.00  & 11.71 & $-6.76$ & 4.59 \\
04498--6842 &   9.18  &   8.03  &  7.51 &   6.62  & 1.27   &  1.19
& \llap{$-$}0.08  & 10.27 & $-8.20$ & 5.17 \\
04539--6821 & (18.93) &  15.50  & 12.70 &   9.70  & 0.18   &  0.21
& \llap{$-$}0.06  & 13.25 & $-5.22$ & 3.98 \\
04557--6753 & (16.46) &  13.66  & 11.56 &   9.10  & 0.41   &  0.34
& \llap{$-$}0.48  & 12.47 & $-6.00$ & 4.29 \\
05003--6712 &  14.20  &  12.40  & 10.80 &   9.20  & 0.17   &  0.17
&   1.04  & 13.09 & $-5.38$ & 4.04 \\
05009--6616 &  15.00  &  12.80  & 11.10 &   8.70  & 0.31   &  0.26
& \llap{$-$}0.19  & 12.59 & $-5.88$ & 4.24 \\
05099--6740 &  14.49  &  12.61  & 11.41 &  10.10  & \llap{$<$}0.07  &
\llap{$<$}0.16 &   \llap{$>$}1.1 & 13.81 & $-4.66$ & 3.75 \\
05112--6755 & (17.15) &  14.30  & 12.30 &   8.90  & 0.41   &  0.33
& \llap{$-$}0.01  & 12.46 & $-6.01$ & 4.29 \\
05112--6739 & (19.13) &  15.70  & 12.90 &   9.80  & 0.31   &  0.16
& 0.08  & 12.91 & $-5.56$ & 4.11 \\
05117--6654 & (15.15) &  12.80  & 11.39 &   9.00  & 0.25   & (0.25)
& \llap{$-$}0.57  & 12.80 & $-5.67$ & 4.16 \\
05128--6455 &  13.50  &  11.80  & 10.36 &   8.40  & 0.41   &  0.60
& \llap{$-$}1.09  & 12.16 & $-6.31$ & 4.41 \\
05190--6748 & (18.84) & (15.25) & 12.40 &   9.00  & 0.44   &  0.30
& \llap{$-$}0.27  & 12.44 & $-6.03$ & 4.30 \\
05203--6638 &  14.50  &  12.10  & 10.50 &   8.70  & 0.15   & (0.15)
& 0.05  & 12.98 & $-5.49$ & 4.08 \\
05291--6700 &  13.10  &  11.10  & 10.10 &   8.80  & \llap{$<$}0.06  &
\llap{$<$}0.19 & \llap{$>$}0.14 & 13.11 & $-5.36$ & 4.03 \\
05329--6709 &  15.70  &  11.60  &  9.40 &   7.30  & 1.27   &  2.73
& \llap{$-$}0.44  & 10.90 & $-7.57$ & 4.92 \\
05348--7024 & (25.10) &  19.30  & 14.40 &  (8.65) & 0.33   &  0.19
&   0.41  & 12.64 & $-5.83$ & 4.22 \\
05360--6648 &  19.70  &  16.60  & 13.90 & (10.60) & 0.12   &  0.12
&   0.70  & 13.84 & $-4.63$ & 3.74 \\
05506--7053 & (17.60) &  14.50  & 12.20 &   8.80  & \llap{$<$}0.13  &
\llap{$<$}0.14 &   \llap{$>$}1.7 & 13.24 & $-5.23$ & 3.98 \\
\hline
\end{tabular}
\end{table*}

The stars can make excursions through colour-colour diagrams due to
their intrinsic variability. For five stars we obtained $N$-band
photometry in 1993 December (paper~II): IRAS01074--7140,
IRAS04407--7000, IRAS04498--6842, IRAS05112--6755, and
IRAS05329--6709 (all probably oxygen stars). We can now compare these
near- and mid-IR magnitudes and colours with those measured in 1994
November (Table 5). The $K$ and $(H-K)$ for 1993 December have been
estimated in the same way as for 1994 November (see above).

The mean difference (``1994 -- 1993'') of the $N$-band magnitudes of
the five stars is $-0.2 \pm 0.2$ mag; there is no indication for a
large systematic difference in the N-band measurements at the two
epochs. Assuming a sinusoidal light-curve, the difference between the
measurements at two random epochs is expected to be a fraction,
1/$\pi$, of the peak to peak amplitude. We assume that the five
difference measurements sample the light curve of a typical
mass-losing oxygen-rich AGB star. This seems reasonable, considering
the time span of nearly a year between the two epochs of measurement,
and typical periods of variability of the order of one to four years.
We thus find the mean peak to peak amplitude (Table 5).

The variability near 10~$\mu$m can also be estimated from a comparison
between the $N$-band measurements (from 1994 November) and the IRAS
fluxes. The latter do not represent the mean fluxes of the stars,
neither are they entirely single-epoch. This is, however, not
important when estimating the typical amplitude of variability, since
the IRAS epoch is certainly very different from the $N$-band epoch. In
Table 6 we list the magnitude differences between the 12~$\mu$m
magnitudes derived from our $N$-band measurements and those derived
from the IRAS data. The mean absolute value of these differences for
the LMC stars is 0.43 mag. Hence the inferred typical amplitude of
variability at 12~$\mu$m is 1.4 mag, which is in excellent agreement
with the value we estimated from the $N$-band data in Table 5.

The $K$- and $N$-band magnitudes vary considerably, while the $(H-K)$
colours do not vary much. The $(K-\left[12\right])$ colours vary
significantly, but not more than do the $K$- or $N$-band magnitudes.
Thus the $K$- and $N$-band magnitudes vary neither in phase, nor in
anti-phase. The phase difference can be estimated by simple vector
calculus on the amplitudes of the $K$- and $N$-band magnitudes and
$(K-\left[12\right])$ colours. In this way we estimate that either of
the $N$- or $K$-band magnitudes follows the other with a phase-lag of
about $\pi/4$, corresponding to a time-lag of 1/8 period. Although
based on five measurements only, the $(K-\left[12\right])$
differences between the two epochs correlate very well with the mean
$(H-K)$ colours: the $(K-\left[12\right])$ colour becomes less
variable as the CSE becomes optically thicker (larger $(H-K)$).

\subsection{Bolometric luminosities}

We have re-determined the bolometric luminosities for two reasons. 
First, the $N$-band measurements are generally of higher accuracy than
the IRAS 12~$\mu$m flux densities. Secondly, we wish to determine the
bolometric luminosities for the particular epoch of our $N$-band
measurements. When IRAS flux densities and NIR magnitudes of large
amplitude variables are obtained at different epochs their combination
may result in an erroneous bolometric luminosity.

The NIR magnitudes have been estimated as described above. The
12~$\mu$m flux densities have been determined from the $N$-band
magnitudes, using $(N-\left[12\right]) = 0.33$ mag. The 25~$\mu$m flux
densities have been derived from these new 12~$\mu$m flux densities,
assuming the same ratios of 25 to 12~$\mu$m flux densities as observed
by IRAS (paper~II). Values in brackets have been estimated from the
relations between the appropriate $(J-H)$, $(K-L)$,
$(\left[12\right]-\left[25\right])$, and $(H-K)$ colours (see
Appendix), where $(\left[12\right]-\left[25\right]) = -2.5 \left[
\log(S_{12}/S_{25}) + \log(6.73/28.3) \right]$ (IRAS Explanatory
Supplement 1988).

%
% FIGURE 5
%
\begin{figure}[tb]
\centerline{\psfig{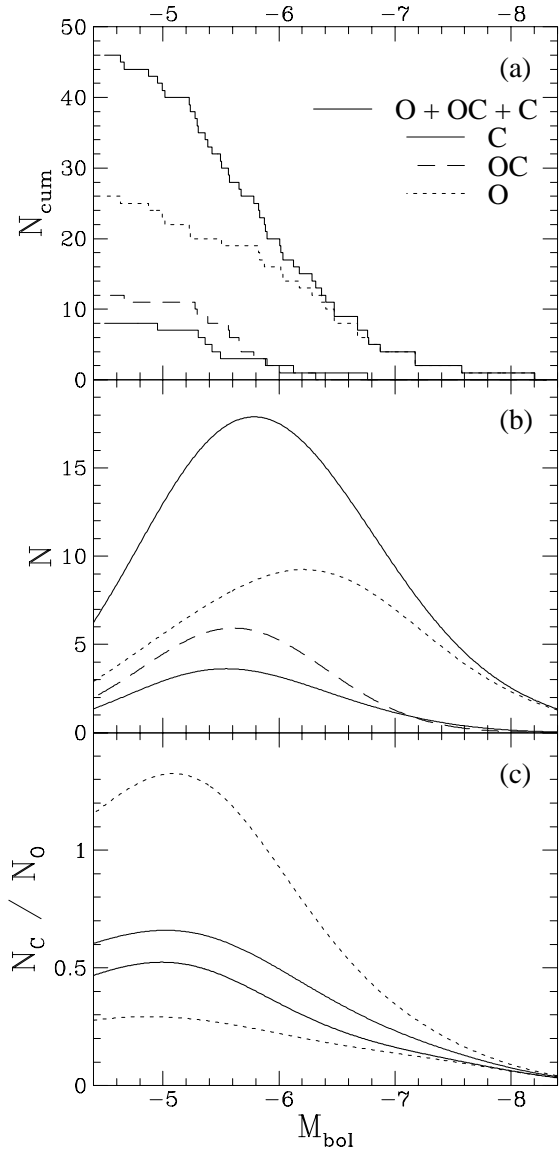}}
\caption[]{(a) Cumulative distribution function of the full sample of
IRAS detected AGB stars in the LMC over absolute bolometric magnitude:
``oxygen'' stars (dotted), ``oxygen or carbon'' stars (dashed),
``carbon'' stars (solid), and their sum (bold solid). (b) Applying
Gaussian broadening (see text), we derive the distribution functions
$N$. (c) From this we derive the distribution function of the ratio of
carbon to oxygen stars for the two extreme cases of carbon star
favoured (all OC stars are C stars) and oxygen star favoured (all OC
stars are O stars), indicated by the dotted line. We also plot the
ratio derived without assuming anything about the OC stars (bold
solid), and assuming half of the OC stars are C stars (thin solid)}
\end{figure}

Bolometric luminosities were obtained by integrating under a
spline-curve as described in paper~II (with photometric zero-points
from Glass \& Feast 1973). We adopted distance moduli of $(m-M)_0 =
18.47$ mag for the LMC and $(m-M)_0 = 18.78$ mag for the SMC (Feast
and Walker 1987). Both lower and higher values for the distance moduli
have appeared in the recent literature (e.g.\ Crotts et al.\ 1995;
Gallagher et al.\ 1996; Sonneborn et al.\ 1997; Feast 1997; Caputo
1997; van Leeuwen et al.\ 1997; Feast \& Catchpole 1997), and from
this we estimate the distance modulus to be known to about $\pm0.15$
mag. The accuracies of the individual bolometric luminosities are
estimated to be typically 0.1 mag, plus the systematic error in the
assumed distance moduli. The contribution to the uncertainty in the
bolometric luminosity from the sometimes poorly known 25~$\mu$m flux
density is less than 0.1 mag. The bolometric luminosity is also
expressed in terms of solar luminosity, where we adopt for the solar
absolute bolometric magnitude $M_{{\rm bol,}\sun} = 4.72$ mag (e.g.\
Sterken \& Manfroid 1992). The results are enumerated in Table 6.

\subsection{Luminosity functions}

We combined the mass-losing AGB stars from the present sample with
the remaining AGB stars from papers II \& III. IRAS05283--6723
(TRM045) was omitted because of its uncertain identification. For
IRAS05003--6712 we took the data quoted here, not that from paper~III,
because now we have simultaneous near- and mid-IR photometry. The
chemical types of the remaining sources from paper~II are estimated
from the $(K-\left[12\right])$ and $(H-K)$ colours, or from additional
data mentioned in paper~II. All of the remaining paper~II sources are
found to be oxygen rich, except for IRAS00554--7351 which is a
confirmed carbon star (Wood et al.\ 1992), and IRAS05295--7121 which
could be either oxygen- or carbon-rich.

The cumulative distribution function of the different types of IRAS
detected AGB stars in the LMC over absolute bolometric magnitude is
presented in Fig.\ 5a. We apply Gaussian broadening by replacing each
star with absolute bolometric magnitude $M_{{\rm bol},\star}$ by
\begin{equation}
n = n_{\rm norm} \times \exp \{-(M_{\rm bol} - M_{\rm bol,\star})^2\}
\end{equation}
where $n_{\rm norm} = 0.564$ mag, to normalise to unity per star. The
FWHM of the star has thereby become 0.83 mag. This is justified by 
both the variability of the sources and the small number of sources in
our sample, and is necessary for deriving the distribution function
over absolute bolometric magnitude (Fig.\ 5b).

From the distribution functions, we derived the dependence of the
ratio of carbon to oxygen stars, on the absolute bolometric magnitude
(Fig.\ 5c). Two extreme cases are plotted (dotted): the carbon-favoured
case, in which all stars classified as ``oxygen or carbon'' stars turn
out to be carbon rich, and the oxygen-favoured case, in which they are
all oxygen rich. We may expect that the category of OC stars is
composed of a mixture of both carbon and oxygen stars. We therefore
also plot the trend that follows from assuming that half of the OC
stars are oxygen stars and the other half carbon stars (solid), as well
as the trend that follows from the probable oxygen and carbon stars
only (bold solid).

\section{Discussion}

\subsection{Completeness and selection effects}

Reid et al.\ (1990) used $V$- and $I$-band imaging to search for stellar
counterparts of IRAS sources in the LMC. Many of these stars, however,
are undetectable at wavelengths below $\sim1 \mu$m. We searched for
IRAS sources at NIR wavelengths, and found $\sim20$ mass-losing AGB
stars more luminous than $M_{\rm bol}\sim-6$ mag. Are there more?

%
% FIGURE 6
%
\begin{figure}[tb]
\centerline{\psfig{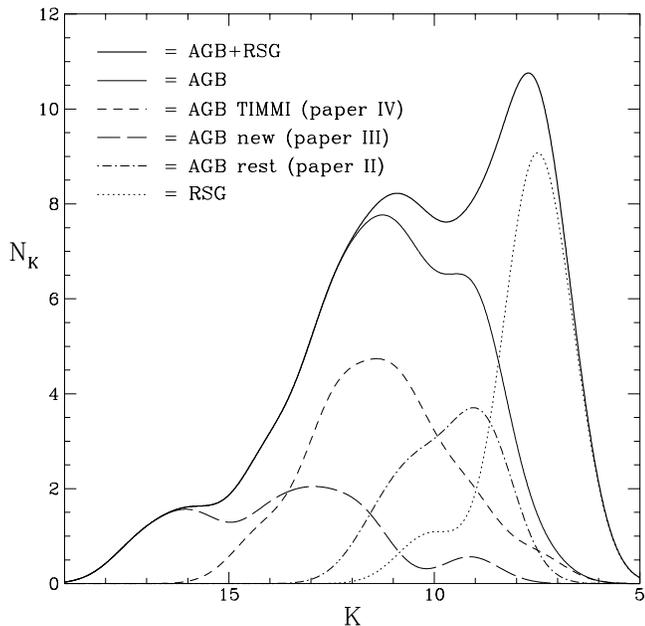}}
\caption[]{Distribution functions $N_K$ over apparent $K$ magnitude, of
IRAS detected LMC stars that have been identified with AGB stars and
RSGs (see text): all known at present (bold solid), all AGB stars known
at present (thin solid), TIMMI sample of AGB stars presented in this
work (short dashed), new AGB stars presented in paper III (long dashed),
remainder of presently known AGB stars (dot-dashed; see paper II), and
all RSGs known at present (dotted)}
\end{figure}

In Fig.\ 6 we plot the distribution functions over apparent $K$-band
magnitude, of stars that have been detected by IRAS and that subsequent
searches have revealed to be AGB stars or RSGs in the LMC. We applied
Gaussian broadening as before. Stars found in $I$-band searches are
generally optically thin in the $K$-band, and the $K$-band magnitude
directly reflects the bolometric magnitude (e.g.\ Wood et al.\ 1992).
But many IRAS detected stars are optically thick in the $K$-band, and
can become as faint as $K > 15$ mag for $M_{\rm bol} \sim -5$ mag. Even
RSGs are found at $K > 10$ mag.

Frogel \& Richer (1983) searched for obscured stars down to a
completeness limit of $K = 11$ mag, and according to Fig.\ 6 they would
have missed a significant fraction of the obscured AGB stars. Frogel et
al.\ (1990) searched for AGB stars in MC clusters down to a limiting
magnitude varying between $K = 12$ and 13. Wood et al.\ (1992) searched
for IRAS counterparts down to a limiting magnitude of $K = 12.5$.
Neither of them are complete for obscured AGB stars. Most of the newly
found obscured AGB stars in paper~III have $K > 12.5$ mag, and we
cannot exclude the possible existence of IRAS detected AGB stars in the
LMC at $K > 17$ mag. It is unlikely that these stars have $M_{\rm bol}
< -6$ mag. There are still some 200 IRAS point sources, however, which
are candidate mass-losing AGB stars in the LMC (paper~I), but which
have not been adequately imaged at NIR wavelengths. These may include
stars with $M_{\rm bol} < -6$ mag. Hence the apparent lack of luminous
mass-losing AGB stars in the LMC may well be a result of
incompleteness.

Selection effects could have been introduced by the detection limits
in the IRAS 12 and 25~$\mu$m bands, and by the detection limits of
the ground-based instruments that were used to search for counterparts
--- mainly at NIR wavelengths. These effects are difficult to
quantify, since the IR properties of obscured carbon and oxygen stars
are not yet known as a function of their bolometric luminosities. The
completeness of our sample of obscured AGB stars is essentially
limited by the IRAS sensitivity at 25~$\mu$m (paper II), i.e.\ it is
probably not affected much by the contribution of the 10~$\mu$m SiO
feature in oxygen-rich CSEs to the IRAS 12~$\mu$m flux-density. The
emissivities of carbon- and oxygen-rich dust differ in such a way that
a carbon star, with the same luminosity and mass-loss rate as an
oxygen-rich star, will be brighter at 25 $\mu$m and fainter in the NIR
than its oxygen-rich counterpart. Thus IRAS may have detected
carbon-rich stars to slightly lower luminosities than oxygen-rich
ones, at 25 $\mu$m. However, the same effect will render our NIR
searches slightly less sensitive to carbon stars than to oxygen stars. 
At present it is not clear what the resulting selection effect is. We
feel that selection effects may become important at $M_{\rm bol} >
-5$ mag.

\subsection{Mid-IR variability}

The amplitude of the $N$-band variations is surprisingly large: 1.4
mag (peak-to-peak). Harvey et al.\ (1974) found a mean 10~$\mu$m
amplitude of $0.88 \pm 0.13$ mag for a sample of galactic OH/IR stars
and Le Bertre (1993) found a mean $N$-band amplitude of $1.01 \pm
0.13$ mag for a sample of galactic mass-losing oxygen-stars. Le
Bertre (1992) found a smaller mean $N$-band amplitude of $0.61 \pm
0.10$ mag for a similar sample of carbon stars.

Harvey et al.\ (1974) found a mean $K$-band amplitude of $1.13 \pm
0.16$ mag for the galactic OH/IR stars that they monitored at
10~$\mu$m. Le Bertre (1992, 1993) found a mean $K$-band amplitude of
$1.08 \pm 0.14$ mag for the galactic mass-losing oxygen stars and a
similar mean $K$-band amplitude of $1.13 \pm 0.16$ mag for galactic
mass-losing carbon-stars, that he monitored in the $N$-band. This is
comparable to the stars for which we had two $N$-band measurements
and that have a $K$-band amplitude of 1.1 mag.

Le Bertre (1988, 1992, 1993) concludes that the mid-IR variability
is due to changes of the bolometric luminosity. The 10~$\mu$m
amplitude of oxygen stars has an additional component due to 
variations of the SiO dust feature. The LMC stars with two $N$-band
measurements have the 10~$\mu$m SiO feature in emission, whereas
galactic OH/IR stars show it in (self-)absorption. As an emission
feature is more sensitive to changes in optical depth and luminosity
than a (self-)absorbed feature, this may explain the large $N$-band
amplitude of the oxygen-rich mass-losing LMC stars compared to their
galactic counterparts --- especially the galactic OH/IR stars.

Our data suggests a phase lag of $\sim 0.13$ pulsation period
between the $K$- and $N$-band light-curves. This seems to be at odds
with the findings for the galactic obscured LPVs that the IR
light-curves between $\sim 1$ and 12~$\mu$m are all in phase with
each other (e.g.\ Nyman \& Olofsson 1986; Jewell et al.\ 1991),
lagging the optical light-curve by between $\sim 0.1$ and 0.2 times
the period. The discrepancy may be explained by the fact that the
stars we investigated do not have extremely optically-thick CSEs.
The observed phase-lag between the $K$- and $N$-band light curves
may reflect a correlation between the emission strength of the
10~$\mu$m SiO feature, and the optical radiation field. We find a
clear indication that the variability of $(K-\left[12\right])$
decreases with increasing $(K-\left[12\right])$ colour, i.e.\ with
increasing optical depth of the CSE. Hence the phase lag between
the $K$- and $N$-band light-curves may disappear in the most
extremely obscured sources, where the 10~$\mu$m feature turns into
(self-)absorption.

\subsection{Frequency of carbon stars}

For the sample of IRAS detected AGB stars in the LMC the ratio of the
number of carbon stars to oxygen stars is in the range 0.21 to 0.83
depending on the number of carbon stars among the OC stars. If the
lower value is correct, the ratio depends only weakly on luminosity,
whereas the higher value would imply a ratio that steeply decreases
with increasing luminosity. The probability that an obscured AGB star
is a carbon star most likely decreases from $\sim 40$\% at $M_{\rm
bol} = -5$ mag ($\log(L/L_\odot) = 3.9$), to $\sim 20$\% at $M_{\rm
bol} = -7$ mag ($\log(L/L_\odot) = 4.7$). At the lower luminosity end
($\log(L/L_\odot) = 3.9$) the fraction of carbon stars is constrained
to be between $\sim22$ and 57\%. At the other extreme, carbon stars
are not completely absent amongst the most luminous AGB stars.

How does this result compare with the luminosity distributions of
optically visible AGB stars in the LMC? Groenewegen \& de Jong
(1993) used a synthetic evolution model to reproduce the observed
optically visible carbon star luminosity function (Cohen et al.\
1981; Richer et al.\ 1979; see also Costa \& Frogel 1996) as well as
the number ratio between carbon and oxygen stars (Blanco \& McCarthy
1983). They find a mean ratio of the number of carbon stars to the
number of oxygen stars of 0.85, in good agreement with the observed
range of 0.6 to 2.2 for the optically visible AGB stars. Amongst the
sample of obscured AGB stars in the LMC carbon stars are less common
than they are amongst optically visible AGB stars. The number of
optically visible AGB stars steeply declines between $M_{\rm bol} =
-5$ and $-6$ mag. The obscured AGB star distribution peaks near
$M_{\rm bol} = -5.8$ mag, and has a gradual high-luminosity drop-off.
Thus the fraction of obscured stars amongst the AGB stars increases
with bolometric luminosity. This is to be expected when optically
visible AGB stars still evolve along the AGB, thereby significantly
increasing in luminosity before becoming dust-enshrouded.

\subsection{The most luminous AGB stars}

The Chandrasekhar limit to the stellar core mass of $1.4 M_{\odot}$
corresponds, via the core mass-luminosity relation, to an AGB
luminosity limit of $M_{\rm bol} \sim -7$ mag (Paczy\'{n}ski 1971).
However, in recent years it has been suggested that for some
intermediate-mass stars the core mass-luminosity relation may not be
valid near the tip of the AGB, if AGB stars become more luminous as
a result of HBB (e.g.\ Bl\"{o}cker \& Sch\"{o}nberner 1991;
Bl\"{o}cker 1995; Marigo et al.\ 1996). On the other hand, Reid et
al.\ (1990) argue that, as a result of intense mass loss on the upper
AGB, stars may not actually become much more luminous than $M_{\rm
bol} \sim -6$ mag.

We discovered that with $M_{\rm bol} \sim -6.8$ mag, the LMC star
IRAS04496--6958 is the most luminous carbon star known in the MCs.
Other luminous mass-losing carbon stars in the MCs are IRAS06028--6722
(LI--LMC1817) with $M_{\rm bol} \sim -6.1$ mag in the LMC (paper~III),
and IRAS00554--7351 with $M_{\rm bol} \sim -6.0$ mag in the SMC
(Whitelock et al.\ 1989; Wood et al.\ 1992). IRAS00350--7436 in the
SMC may also be a carbon star: Whitelock et al.\ (1989) found that it
has a K-type spectrum with weak C$_2$ absorption bands, and they
explain this object as being either a carbon-rich post-AGB object in
the transition to the planetary nebula phase, or an interacting
binary. We estimate $M_{\rm bol} \sim -6.8$ mag, i.e.\ equally
luminous as IRAS04496--6958. Confirmation that IRAS00350--7436 has
become carbon enriched due to dredge-up is crucial.

The two brightest LMC stars in our TIMMI sample are IRAS04498--6842
and IRAS05329--6709, both probable oxygen stars. In 1994 November they
were very luminous: $M_{\rm bol} \sim -8.2$ and $-7.6$ mag
respectively. For IRAS05329--6709 this may be due to variability:
compare the NIR brightness in 1994 November to its light curve (Wood
et al.\ 1992). The NIR variability of IRAS04498--6842 suggests it is
a Mira variable, while its rather blue $(H-K)$ colour is more
characteristic of a RSG. With $M_{\rm bol} \sim -7.2$ and $-6.7$ mag,
IRAS04509--6922 and IRAS04516--6902 are amongst the most luminous
AGB stars, and our optical/NIR spectra have shown them to be oxygen
rich. IRAS05316--6604 (WOH SG374) was listed in paper~II as an AGB
star candidate with $M_{\rm bol} \sim -7.2$ mag, but its near/mid-IR
colours suggest it may be a RSG.

IRAS00165--7418 (VV Tuc) is probably a RSG, judging from its high
bolometric luminosity ($M_{\rm bol} \sim -9.1$ mag) and its location
in the $(K-\left[12\right])$ versus $(H-K)$ diagram. The oxygen-rich
nature of IRAS01074--7140 ($M_{\rm bol} \sim -7.0$ mag) is in
agreement with its spectral type of M5e (Whitelock et al.\ 1989).

\subsection{Hot Bottom Burning}

It has been proposed that in AGB stars with massive mantles, the
convective layer may reach the nuclear burning shells. This so-called
Hot Bottom Burning (HBB; e.g.\ Boothroyd et al.\ 1995, and references
therein) could prevent stars with progenitor masses $\sim 6$ to $7
M_{\sun}$ and solar metallicity from becoming carbon stars, and may
increase their bolometric luminosities by an unknown factor. For half
solar metallicity --- which is typical for the LMC (Russell \&
Bessell 1989; Russell \& Dopita 1990) --- this may occur in AGB stars
with a progenitor mass as low as $\sim 5 M_{\sun}$.

Reid et al.\ (1988, 1995) did not find any carbon star brighter than
$M_{\rm bol} = -6$ mag amongst their LMC LPVs. However, we find
IRAS04496--6958 to be a very luminous, $M_{\rm bol} = -6.8$ mag,
carbon star. At this luminosity AGB stars are expected to be oxygen
rich due to the effects of HBB. On the other hand, we also find
oxygen-rich AGB stars with luminosities as faint as $M_{\rm bol} \sim
-5$ mag, where mass-losing AGB stars are expected to be carbon rich.

Intense mass loss decreases the mass contained in the stellar mantle,
and HBB may switch off. Hence very luminous AGB stars may become
carbon-rich again, just before they leave the AGB. The possibly very
luminous carbon-rich post-AGB object IRAS00350--7436 (see above) may
be an example of this scenario.

\section{Conclusions}

We have studied the chemical composition of the photospheres and
circumstellar envelopes of the presently known sample of obscured
(i.e.\ IRAS detected) AGB stars in the LMC. Because of their high
mass-loss rates these stars are expected to be representative of AGB
stars upon termination of their evolution along the AGB. Thus they
are different from optically visible AGB stars, which may still evolve
in luminosity and photospheric chemical composition before leaving the
AGB. The bolometric luminosity should therefore be a much better
measure of the main-sequence mass for an obscured AGB star, than it is
for an optically visible AGB star.

Our search for new OH maser emitters in the LMC resulted in one
new detection, from IRAS04407--7000. For two stars with moderately
thick CSEs spectra around 800~nm were obtained. For the remainder of
the sample, we used the $(K-\left[12\right])$ versus $(H-K)$ diagram
to establish the probable chemical composition of their CSEs. The
diagram may also distinguish between RSGs and AGB stars: RSGs have
optically thinner CSEs (bluer $(H-K)$ colours) than do AGB stars, at
the same 12~$\mu$m excess with respect to the $K$-band magnitude. This
is because RSGs are bigger than AGB stars, and so are the inner radii
of their respective CSEs.

The ratio of the number of carbon stars to the number of oxygen
stars is found to decrease with increasing luminosity. Carbon stars
are very rare at luminosities exceeding $M_{\rm bol} \sim -6$ mag,
but not absent. The most luminous carbon star in the MCs is found to
be IRAS04496--6958, with $M_{\rm bol} = -6.8$ mag. Yet there are still
many mass-losing stars at luminosities about $M_{\rm bol} \sim -5$ mag
that are oxygen rich. It is puzzling to find carbon stars and oxygen
stars at the same luminosity at the tip of the thermally pulsing AGB.
Possibly the sample of mass-losing AGB stars represents a range of
metallicities (cf.\ Olszewski et al.\ 1996), in which case the
luminous carbon stars would be metal rich and therefore not experience
HBB, whereas the less luminous oxygen stars would be metal poor and
therefore already experience HBB at low luminosities.

\acknowledgements{We would like to thank Peter Wood for providing us
with data on his OH/IR and SHV stars, Hans-Ulrich K\"{a}ufl for helping
with the observations of the TIMMI standard stars, Jay Frogel for
critically reading the manuscript, and the referee Prof.\ Harm Habing
for his comments which helped improve the paper. We acknowledge the
granting of Director's Discretionary Time for obtaining the NTT spectra.
Jacco lamenta de estar deprivado de la mejor manera para agradecer a
Montse como ella lo merecio mas que nadie.}

\appendix

\section{TIMMI standard stars}

%
% TABLE A1
%
\begin{table*}[tb]
\caption[]{Magnitudes of TIMMI standard stars: Bright Star Catalogue
entries, names, spectral types, magnitude band ($N_0$, $N_1$, $N_2$,
or $N_3$), and magnitudes: Koornneef \& Yagnam (ESO, 1985,
unpublished), van der Bliek et al.\ (1996), and our observations, with
their 1-$\sigma$ error estimates (where known). We adopted the
magnitudes from van der Bliek et al.\ for $\alpha$~CMa ($N_0$) and
$\alpha$~Car ($N_{1,2,3}$).}
\begin{tabular}{lllllll}
\hline\hline
BSC entry & Name & Spectral type & magnitude band & Koornneef \&
Yagnam & van der Bliek et al.\ & this paper \\
\hline
HR 1264 & $\gamma$  Ret  & M4 {\sc iii}          & $N_0$ & --0.71 &
--0.73 $\pm$ 0.03 & \llap{$-$}0.87 $\pm$ 0.03 \\
        &                &                       & $N_1$ & --0.65 &
--0.64 $\pm$ 0.02 & \llap{$-$}0.75 $\pm$ 0.04 \\
        &                &                       & $N_2$ & --0.73 &
--0.74 $\pm$ 0.01 & \llap{$-$}0.85 $\pm$ 0.04 \\
        &                &                       & $N_3$ & --0.80 &
--0.82 $\pm$ 0.02 & \llap{$-$}0.93 $\pm$ 0.04 \\
HR 2326 & $\alpha$  Car  & F0 {\sc ii}           & $N_0$ & --1.52 &
--1.46 $\pm$ 0.02 & \llap{$-$}1.44 $\pm$ 0.02 \\
        &                &                       & $N_1$ & --1.51 &
--1.45 $\pm$ 0.02 & \ \ \ \ adopted   \\
        &                &                       & $N_2$ & --1.52 &
--1.47 $\pm$ 0.02 & \ \ \ \ adopted   \\
        &                &                       & $N_3$ & --1.45 &
--1.44 $\pm$ 0.02 & \ \ \ \ adopted   \\
HR 2491 & $\alpha$  CMa  & A1 {\sc v}            & $N_0$ & --1.41 &
--1.40 $\pm$ 0.02 & \ \ \ \ adopted   \\
HR 3634 & $\lambda$ Vel  & K4 {\sc i}b-{\sc ii}  & $N_0$ & --1.78 &
--1.77 $\pm$ 0.02 & \llap{$-$}1.76 $\pm$ 0.02 \\
HR 3748 & $\alpha$  Hya  & K3 {\sc ii}-{\sc iii} & $N_0$ & --1.30 &
--1.30 $\pm$ 0.01 & \llap{$-$}1.39 $\pm$ 0.02 \\
HR 3903 & $\upsilon$ Hya & G7 {\sc iii}(b)       & $N_0$ &        &
                  & 1.74 $\pm$ 0.02 \\
HR 4174 & $\gamma$ Cha   & M0 {\sc iii}          & $N_0$ &        &
                  & \llap{$-$}0.14 $\pm$ 0.04 \\
\hline
\end{tabular}
\end{table*}

On 1995 December 4 and 5 we measured several standard stars, in
order to examine the quality of the currently available standard-star
magnitudes. The results are presented in Table 7 for the broad-
($N=N_0$) and narrow-band ($N_{1,2,3}$) magnitudes. The $N_1$, $N_2$,
and $N_3$ filters are centred at 8.4, 9.8, and 12.6~$\mu$m,
respectively (http://www.eso.org/proposals/timmi.html). We quote the
magnitudes that have been in use at ESO from a compilation by
Koornneef \& Yagnam (ESO, 1985, unpublished) as well as more recent
measurements, obtained with the ESO InSb bolometer by van der Bliek
et al.\ (1996). The bright stars $\alpha$~CMa (Sirius) and
$\alpha$~Car (Canopus) were adopted as standards for the broad- and 
narrow-band magnitudes respectively; using the magnitudes given by
van der Bliek et al.\ (1996). The stars $\gamma$~Ret and
$\lambda$~Vel are suspected by Koornneef \& Yagnam to be variable.
This would not be surprising for a giant of a spectral type as late
as M4 ($\gamma$~Ret) or a late-type supergiant ($\lambda$~Vel).

By studying the airmass dependence of the differences between our
photometry and that by van der Bliek et al.\ (1996) we estimated
the atmospheric extinction coefficients. For the $N_0$-band these
coefficients were 0.50 and 0.46 mag/airmass for the first and second
night respectively. For the $N_1$, $N_2$, and $N_3$-bands on the
second night the coefficients were 0.36, 0.38, and 0.40 mag/airmass
respectively.

Our observed magnitude for $\lambda$~Vel is in excellent agreement
with those found by the above mentioned authors, although it is a
suspected variable. Late-type supergiants are generally less
variable than are AGB stars. There is less agreement about the
magnitudes of $\alpha$~Car, $\alpha$~Hya (Alphard) and especially
$\gamma$~Ret. It is not likely that the discrepancy for
$\gamma$~Ret has to do with the differences in the spectral
response functions of the bolometer and the camera: $\lambda$~Vel
and $\gamma$~Ret have similar spectral slopes within the $N$-band,
as seen from their $N_1$-, $N_2$-, and $N_3$-band magnitudes (Table
7). It is more likely that $\gamma$~Ret is indeed variable. Our
$N_0$-band magnitude of $\alpha$~Car confirms the result by van der
Bliek et al.\ (1996). Either this star is variable, or the magnitude
listed by Koornneef \& Yagnam is incorrect. Van der Bliek et al.\
(1996) do agree with Koornneef \& Yagnam on the $N_0$-band magnitude
of $\alpha$~Hya, whereas we measure it to be brighter by 0.09 mag.
The SAAO NIR standard stars $\upsilon$~Hya and $\gamma$~Cha (Carter
\& Meadows 1995) have been observed for the first time in the
$N$-band, and they may be suitable for use as standard stars if
they are confirmed to be non-variable. Both are fainter than the
standards currently in use.

We conclude that it is difficult to obtain photometric accuracies
better than $\sim$5 or 10\% in the $N$-band with TIMMI, as the
accuracy is limited by the reliability of the available standard
star values. The error in an individual measurement for a star with
$N$ between $\sim3$ and $\sim4$ mag is dominated by the uncertainty
in the adopted magnitude of the standard star used. Measurements of
stars fainter than $N\sim6$ mag are dominated by the intrinsic noise
in the measurement. Together with uncertainties involved in the
extinction correction, this combines with the internal noise on an
individual measurement into the 1-$\sigma$ error estimates given in
Table 1.

\section{IR colour relations}

We have investigated the IR colour relations for the $J$, $H$, $K$,
and $L$-bands in the SAAO photometric system, and the IRAS 12 and
25~$\mu$m bands. We define $\left[12\right] = -2.5 \log(S_{12}/28.3)$
and $\left[25\right] = -2.5 \log(S_{25}/6.73)$, where $S_{12}$ and
$S_{25}$ are the flux densities in Jy at 12 and 25~$\mu$m, respectively
(IRAS Explanatory Supplement 1988). The SAAO $L$-band has an effective
wavelength of 3.45~$\mu$m.

%
% FIGURE 7
%
\begin{figure}[tb]
\centerline{\psfig{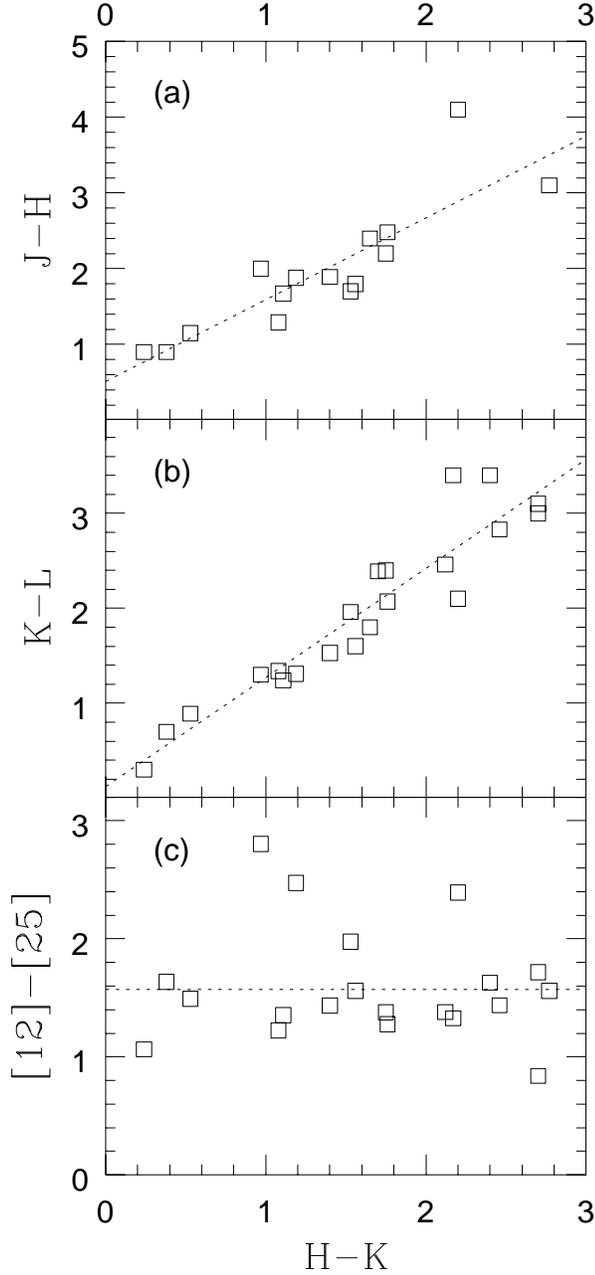}}
\caption[]{$(J-H)$ (a), $(K-L)$ (b), and
$(\left[12\right]-\left[25\right])$ (c) colours versus $(H-K)$ colour
for the stars of our sample. The dotted lines are linear fits to the
data}
\end{figure}

We plot the $(J-H)$, $(K-L)$, and $(\left[12\right]-\left[25\right])$
colours versus the $(H-K)$ colour (Fig.\ 7). The dotted lines are linear
fits to the data points:
\begin{eqnarray}
\left( \begin{array}{c} (J-H) \\ (K-L) \\
(\left[12\right]-\left[25\right]) \end{array} \right) = & \left(
\begin{array}{c} 0.51 \pm 0.43 \\ 0.12 \pm 0.30 \\ 1.57 \pm 0.49
\end{array} \right) & + \nonumber \\  & \left( \begin{array}{c} 1.08
\pm 0.17 \\ 1.15 \pm 0.09 \\ 0 \end{array} \right) & \times (H-K)
\end{eqnarray}

Cohen et al.\ (1981) showed that the $(H-K)$ and $(J-H)$ colours of
optically visible carbon stars in the Milky Way are correlated, from
which we derive on the SAAO system (Carter 1990):
\begin{equation}
(J-K) = 0.76 + 1.95 \times (H-K)
\end{equation}
They explained the correlation as a result of line blanketing by
molecular absorption bands. From Costa \& Frogel (1996) we derive a
correlation between their $(J-K)$ and $(H-K)$ colours, transformed to
the SAAO system (Carter 1990), for optically visible carbon stars in
the LMC:
\begin{equation}
(J-K) = 0.77 + 1.83 \times (H-K)
\end{equation}
The selected sample of Milky Way stars from Guglielmo et al.\ (1993)
yield, after transformation to the SAAO system (Carter 1990), for the
obscured carbon stars:
\begin{equation}
(J-K) = (0.81 \pm 0.29) + (2.29 \pm 0.03) \times (H-K)
\end{equation}
and for the obscured oxygen stars:
\begin{equation}
(J-K) = (0.70 \pm 0.26) + (2.43 \pm 0.03) \times (H-K)
\end{equation}
For the obscured AGB stars in the LMC we derive:
\begin{equation}
(J-K) = (0.51 \pm 0.43) + (2.08 \pm 0.17) \times (H-K)
\end{equation}

%
% FIGURE 8
%
\begin{figure}[tb]
\centerline{\psfig{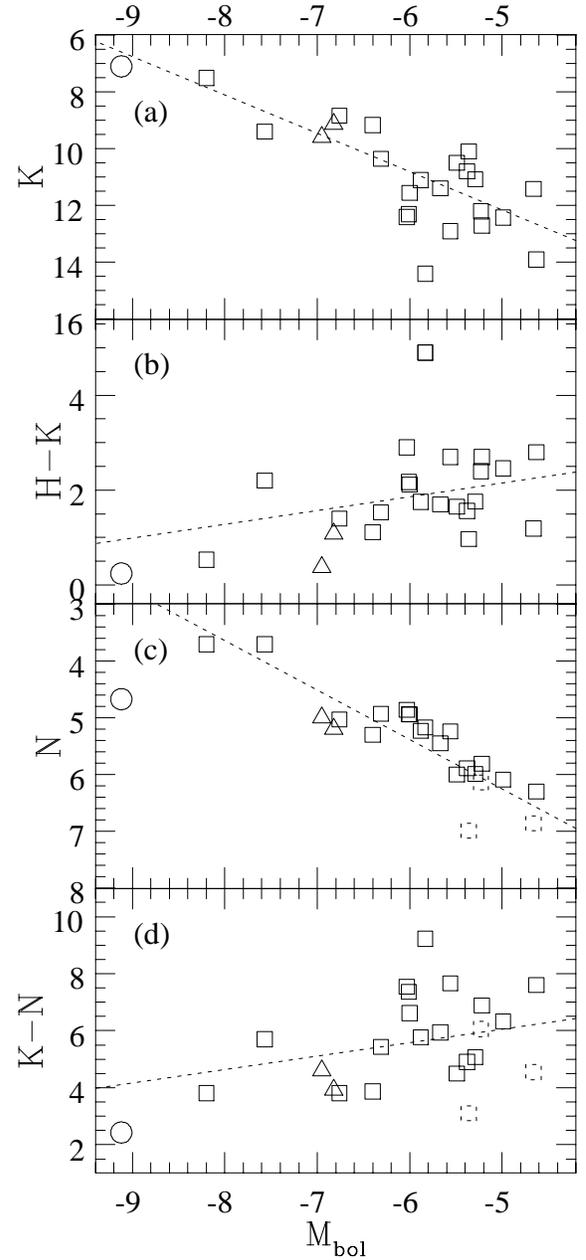}}
\caption[]{$K$ (a), $(H-K)$ (b), $N$ (c), and $(K-\left[12\right])$
(d) as a function of absolute bolometric magnitude for the LMC stars
(squares, dotted in the case of lower limits to the $N$-band magnitudes
and upper limits to the $(K-N)$ colours, respectively), SMC stars
(triangles), and VV Tuc (circle). The dotted lines are linear fits to
the LMC data (see text)}
\end{figure}

We plot $K$ and $N$-band magnitudes and $(H-K)$ and $(K-\left[12\right])$
colours versus the bolometric luminosity (Fig.\ 8). The dotted lines
are linear fits to the LMC data points (we also include the stars with
N-band lower limits, to diminish bias):
\begin{eqnarray}
\left( \begin{array}{c} K \\ (H-K) \\ N \\ (K-N) \end{array} \right) =
\left( \begin{array}{r} 18.9 \pm 1.0 \\ 3.6 \pm 0.6 \\ 10.6 \pm 0.4 \\
8.4 \pm 1.1 \end{array} \right) + \left( \begin{array}{c} 1.35 \pm 0.26
\\ 0.29 \pm 0.16 \\ 0.87 \pm 0.10 \\ 0.47 \pm 0.28 \end{array} \right)
\times M_{\rm bol}
\end{eqnarray}
Bolometrically fainter stars are relatively fainter in the $K$-band,
and relatively brighter in the $N$-band, yielding larger
$(K-\left[12\right])$ colours. Their $(H-K)$ colours are also larger,
indicating that fainter stars are optically thicker: at equal
mass-loss rates, they have smaller inner radii of the CSEs, and
consequently larger dust column densities than the more luminous
stars. Thus fainter stars have fainter $K$-band magnitudes due to
increased circumstellar opacity, and brighter $N$-band magnitudes due
to increased circumstellar dust emission.

\end{document}